\documentclass[twocolumn,trackchanges]{aastex63}

\newcommand{\Cloudy}{\textsc{Cloudy}}

\font\manual=manfnt at 7pt \def\dbend{\hbox{\raise0.9ex\hbox{\manual\char127\hspace{0.6em}}}}

\providecommand{\e}[1]{\ensuremath{\times 10^{#1}}}

\newcounter{INTERNALionstage}

\def\gtsim{\mathrel{\hbox{\rlap{\hbox{\lower4pt\hbox{$\sim$}}}\hbox{$>$}}}}
\def\lesssim{\mathrel{\hbox{\rlap{\hbox{\lower4pt\hbox{$\sim$}}}\hbox{$<$}}}}

\def\A{{\rm\thinspace \AA}}

\def\eV{{\rm\thinspace eV}}
\def\g{{\rm\thinspace g}}

\def\K{{\rm\thinspace K}}
\def\keV{{\rm\thinspace keV}}

\def\m{{\rm\thinspace m}}

\def\s{{\rm\thinspace s}}

\def\W{{\rm\thinspace W}}

\def\izw{\mbox{$\rm\thinspace I\rm\thinspace ZW\rm\thinspace 1 $}}
\def\hi{\mbox{{\rm H~{\sc i}}}}

\def\heii{\mbox{{\rm He~{\sc ii}}}}

\def\ciii{\mbox{{\rm C~{\sc iii}}}}
\def\civ{\mbox{{\rm C~{\sc iv}}}}

\def\mgii{\mbox{{\rm Mg~{\sc ii}}}}

\def\feii{\mbox{{\rm Fe~{\sc ii}}}}

\def\h0{\mbox{{\rm H}$^0$}}
\DeclareMathAlphabet{\vib}{OML}{cmm}{m}{it}

\received{September 6, 2020}
\accepted{November 14, 2020}
\submitjournal{ApJ}

\graphicspath{{./}{figures/}}

\begin{document}


\title{Improved $\feii$ emission line models for AGN using new atomic datasets}

\author[0000-0002-5222-1337]{A. Sarkar}
\affiliation{University of Kentucky, 
505 Rose street,
Lexington,KY 40506, USA}

\author[0000-0003-4503-6333]{G. J. Ferland}
\affiliation{University of Kentucky, 
505 Rose street, 
Lexington,KY 40506, USA}

\author[0000-0002-8823-0606]{M. Chatzikos}
\affiliation{University of Kentucky, 
505 Rose street, 
Lexington,KY 40506, USA}

\author[0000-0002-2915-3612]{F. Guzm\'an}
\affiliation{University of North Georgia, 
Dahlonega, GA 30597, USA}

\author[0000-0001-7490-0739]{P. A. M. van Hoof}
\affiliation{Royal Observatory of Belgium, Ringlaan 3, B-1180 Brussels, Belgium}

\author[0000-0002-4359-1408]{R. T. Smyth}
\affiliation{Centre
for Theoretical Atomic,
Molecular and Optical Physics, 
School of Mathematics and Physics,
Queen’s University Belfast, Belfast BT7 1NN,
Northern Ireland, UK}

\author{C. A. Ramsbottom}
\affiliation{Centre
for Theoretical Atomic,
Molecular and Optical Physics,
School of Mathematics and Physics,
Queen’s University Belfast, Belfast BT7 1NN,
Northern Ireland, UK}

\author[ 	
0000-0001-5435-1170]{F. P. Keenan}
\affiliation{Astrophysics Research Centre,
School of Mathematics and Physics,
Queen’s University Belfast,
Belfast BT7 1NN, Northern Ireland, UK.}

\author{C. P. Ballance}
\affiliation{Centre
for Theoretical Atomic,
Molecular and Optical Physics,  
School of Mathematics and Physics,
Queen’s University Belfast, Belfast BT7 1NN,
Northern Ireland, UK}

\begin{abstract}
Understanding the $\feii$ emission 
from Active Galactic Nuclei (AGN) 
has been a grand challenge for 
many decades. The rewards from 
understanding the AGN spectra would
be immense, involving both quasar 
classification schemes such as ``Eigenvector 
1'' 
and tracing the chemical evolution of the 
cosmos. Recently, three large $\feii$ atomic 
datasets with radiative and electron 
collisional
rates have become available. 
We have incorporated these into the 
spectral synthesis code $\Cloudy$ and examine 
predictions using a new generation of 
AGN Spectral Energy Distribution (SED), which
indicates that 
the UV emission can be quite different 
depending on the dataset utilised.
The Smyth et al dataset better 
reproduces 
the observed $\feii$ template of the $\izw$ 
Seyfert
galaxy in the UV and optical regions, and we 
adopt these data.
We consider both thermal and microturbulent 
clouds  
and show that a microturbulence 
of $\approx$ 100 km/s reproduces the observed 
shape and strength of the so-called $\feii$ 
``UV bump''. 
Comparing our predictions with the 
observed $\feii$ template, we derive a typical
cloud density of $10^{11}$ cm$^{-3}$  and 
photon flux 
of $10^{20}$ cm$^{-2}$ s$^{-1}$, and show that 
these largely 
reproduce the observed $\feii$ 
emission in the UV and optical. 
We calculate the $I(\feii)/I(\mgii)$ 
emission-line intensity ratio 
using our best-fitting model and obtain 
log($I(\feii)/I(\mgii)$) $\sim$ 0.7, suggesting
many AGNs have a 
roughly solar Fe/Mg abundance ratio. 
Finally,  
we vary the Eddington ratio and SED shape as a 
step in understanding 
the Eigenvector 1 correlation. 

\end{abstract}


\section{Introduction} \label{sec:intro}
Emission lines of $\feii$ are a major 
contributor to the AGN spectra, 
with wavelengths spanning the infrared (IR) to 
ultraviolet (UV) regions. 
These emission lines provide  
an important laboratory for 
developing $\feii$ as a physical 
and geometrical
diagnostic for the broad--line 
regions (BLRs) in AGN.
Understanding the physics behind 
the $\feii$ emission is important 
for multiple reasons.
It is the primary 
indicator of Eigenvector 1 (EV1), 
a key property to construct the quasar 
main sequence 
\citep[e.g.,][]{1992ApJS...80..109B,2000ApJ...536L...5S,2001ApJ...558..553M}.
EV1, which relates to the 
black hole mass and quasar 
orientation \citep[e.g.,][]{2014Natur.513..210S,2018ApJ...866..115P,2019ApJ...882...79P}, 
shows a strong 
anti-correlation with optical $\feii$ 
equivalent width in the Seyfert galaxies and 
quasars. Recent work 
shows correlations between the line 
widths of optical $\feii$ and UV $\feii$ 
in quasar spectra, which 
indicates that the emitting 
regions are close together 
\citep{2015ApJS..221...35K}. 
It may be that both 
are emitted from the outer region 
of BLR or intermediate-line region (ILR).
 

Strong $\feii$ emission is also useful 
to investigate the energy budget of 
the emitting gas.
The iron abundance as a function of 
cosmic time allows us to verify 
several cosmological parameters 
\citep{1999ARA&A..37..487H}.
In  most galaxy evolution models, 
iron is mainly deposited in the 
interstellar medium (ISM) through 
Type Ia supernovae (SNIa), which occur 
about 0.3 to 1 billion years after
the initial burst of star formation, 
because for Type Ia supernovae 
to occur 
the galaxy has to be old enough 
to host white dwarfs. 
This triggers a sudden jump in the 
iron abundance, as shown in 
\citet[][]{1999ARA&A..37..487H} 
and \citet{2001ApJ...558..351M}. 
The Gunn-Peterson effect in quasars 
suggests that the 
most recent onset of star formation 
occurred around $z\ \sim$ 6 
\citep[e.g.,][]{2001ApJ...560L...5D,2006ARA&A..44..415F,2007MNRAS.374..493B,2009ApJ...695..809K,2019PASP..131g4101S}.  
A properly calibrated iron chronometer 
would allow us to measure the iron 
abundance 
at high redshift, offering the 
possibility of measuring the redshift 
when 
Type Ia supernovae first happened 
\citep{2004ApJ...615..610B}. 

The timescale for iron enrichment in 
the ISM is much longer than that of 
$\alpha$-elements, such as- magnesium. 
Magnesium is deposited in the 
ISM via core-collapsed 
(Type II) supernovae, which have 
much shorter timescales than Type I. 
The flux ratio of UV $\feii$ 
multiplet to the 
$\mgii\ \lambda$2800 doublet 
(hereafter $I(\feii)/I(\mgii)$) of quasars, 
therefore lets us probe the material 
deposited through 
Type Ia supernovae versus that 
processed through $\alpha$-process in 
stars and then ejected via Type II 
supernovae 
\citep[e.g.,][]{2007ApJ...669...32K,2011ApJ...739...56D,2015Natur.518..512W,2017ApJ...849...91M,2019ApJ...874...22S}. 
Thus, the $I(\feii)/I(\mgii)$ ratio provides a 
powerful approach to investigate the 
chemical evolution of AGNs. 

The theoretical and observational 
aspects of $\feii$ emission have 
been long-standing and important 
problems
addressed by 
\citet{1977ApJ...215..733O, 1978ApJS...38..187P,1983ApJ...275..445N,1985ApJ...288...94W,1999ApJS..120..101V,2004ApJ...615..610B} 
due to its great strength in many
Seyfert galaxies and quasars. 
Previous work with collisional 
excitation of $\feii$ 
in the framework of photoionization 
models  failed to reproduce the 
observed strength of $\feii$ emission 
in the UV. 
The observed $\feii$ spectra of 
narrow-line Seyfert galaxies 
\citep[e.g.,][]{2001ApJS..134....1V} 
contain a so called ``UV bump" 
between the $\ciii\ \lambda$1909 and $\mgii\ \lambda$2800
emission lines \citep{2004ApJ...615..610B},  
which is produced by blending of a 
large number of $\feii$ emission lines
due to transitions between 
high-lying states with energies 
E $\geq$ 13.25 eV \citep[e.g.,][]{2007ApJS..173....1L,2008ApJ...675...83B}. 
However, the collisional excitation 
models \citep{1981ApJ...250..478K} only predict an electron 
temperature of T $\leq$ $10^4$ K 
at the illuminating faces of the 
$\feii$
emitting cloud which is too low 
to excite electrons to the 
high-lying energy levels (E $\geq$ 8 eV).

\citet{1985ApJ...288...94W} first 
pointed out the importance of 
continuum fluorescence to excite 
the electrons 
to the higher energy states 
(E $\geq$ 11.6 eV). Several 
spectral energy distributions 
(SEDs) of AGN with 
increasing Eddington ratio 
(L/L$_{{\rm Edd}}$) have been proposed 
to reproduce the observed shape 
of the $\feii$ UV bump, 
where L$_{{\rm Edd}}$ is the Eddington
luminosity of the corresponding AGN
\citep[e.g.,][]{1987ApJ...323..456M,1997ApJS..108..401K,2012MNRAS.425..907J}. 
However, the $\feii$ atomic 
dataset has long remained
a concern for reproducing 
$\feii$ spectra. A larger $\feii$ 
model involves a large number of 
transitions between 
the high-lying energy states, producing 
stronger $\feii$ emission in the UV. 

We have incorporated three 
recent $\feii$ datasets,
namely those of
\citet{2015ApJ...808..174B}, 
\citet{2018PhRvA..98a2706T}, and 
\citet{2019MNRAS.483..654S} 
into the spectral synthesis code $\Cloudy$ 
\citep{2017RMxAA..53..385F} in addition to the 
\citet{1987ApJ...323..456M}, 
\citet{1997ApJS..108..401K}, and 
\citet{2012MNRAS.425..907J} AGN SEDs. 
Our model predictions are compared with 
the observed UV 
\citep{2001ApJS..134....1V} and 
optical \citep{2004A&A...417..515V} 
BLR templates of
1 ZW I Seyfert galaxy to 
constrain the physical conditions 
and geometrical properties 
of the $\feii$ emitting gas. 
We find that the 
\citet{2019MNRAS.483..654S} 
$\feii$ dataset, the 
\citet{2012MNRAS.425..907J} SED, 
and a dense turbulent cloud,
largely reproduce the Fe II 
emission with solar abundances.

{
Our goal is to  reproduce the 
shape and strength of the $\feii$ 
UV bump
and optical emission.
We study the emission from a single BLR cloud in some detail.
The emission lines are known to be formed in a distribution of clouds 
with different locations and densities.
This was first measured with reverberation \citep{1993PASP..105..247P}
and is a consequence of atomic physics selection effects \citep{1995ApJ...455L.119B}.
This must be taken into account when emission from different species
with a broad range of ionization potential or critical density is considered. 
However, emission from one particular species is generally localized to 
favored values of the density and ionizing photon flux, as illustrated by 
\citet{1997ApJS..108..401K} and \citet{2004ApJ...615..610B}, so, 
when considering a single species like $\feii$, a typical set of cloud parameters can be considered, as in \citet{2009ApJ...707L..82F}.  }
{We compare $\feii$ and $\mgii$ emissions later in the paper because these species
are used as abundance indicators in high-redshift quasars.  
Figure 3f of \citet{1997ApJS..108..401K} 
shows that $\feii$ and $\mgii$ lines form in
very similar clouds, justifying this approach.}

{There are also complex interplays between different sets of model parameters
such as turbulence, the SED shape, or composition. Our primary goal is to develop a
testing framework to compare theory and observations, and to document the spectral
properties of these new atomic data sets.  
The new $\feii$ atomic data files, and the infrastructure needed to use them,
will be included in the C17.03 update to $\Cloudy$ and we hope that the 
analysis methods we demonstrate here can serve as a guide for future studies and
detailed comparisons with observations.  
}

{We assume  solar abundances.
Studies of high-ionization lines find that the metallicity is 
above
solar and correlates with luminosity 
\citep{1999ARA&A..37..487H}.
Recently \citet{2020arXiv201006902S} measured the 
$\feii$/$\mgii$ ratio
of a large number of quasars across cosmic time.  They did not 
find any
evolution and concluded
that the ratios were consistent with solar 
abundances.
In later sections of the paper we show that this line
ratio has only a weak metallicity dependence.
For simplicity, we assume solar abundances in the calculations presented here.
}

{Finally, we adopt a 
cosmology of $H_{0}$ = 70 km s$^{-1}$ 
Mpc$^{-1}$, 
$\Omega_{\Lambda}$ = 0.7, and $\Omega_{m}$ = 
0.3.
}





\section{Basic ingredients}\label{sec:basic}
A complete model of the physical 
processes that affect the $\feii$ 
spectrum would help us to 
constrain the gas properties 
and dynamics in the BLR. 
However, difficulties arise 
because a variety of processes 
determine the observed $\feii$ 
emission, such as collisional 
excitation, pumping by the 
continuum photons, and fluorescence 
via line overlap. 
The spectral energy distribution 
(SED) of the background AGN, the cloud's 
density, 
turbulence, and optical depth of 
the gas also play important roles. 
A realistic model should 
take into account all those 
effects while modeling the 
temperature and ionization 
structure of the gas.

We use the development version of 
\Cloudy\ , last described by 
\citet{2017RMxAA..53..385F}.
We adopt \Cloudy's default solar 
composition, as listed in 
Table \ref{table:solar_abun}.

\begin{deluxetable}{cccc}
\tablenum{1}
\tablecaption{Solar abundances used in $\Cloudy$}
\tablewidth{0pt}
\tablehead{
\colhead{Elements} & \colhead{Abundances} & \colhead{Elements} & \colhead{Abundances}}
\startdata
H & 1.0 & S & 1.84e-5 \\
He & 0.10 & Cl & 1.91e-7 \\
Li & 2.04e-9 & Ar & 2.51e-6 \\
Be & 2.63e-11 & K & 1.32e-7 \\
B & 6.17e-10 & Ca & 2.29e-6 \\
C & 2.45e-4 & Sc & 1.48e-9 \\
N & 8.51e-5 & Ti & 1.05e-7 \\
O & 4.90e-4 & V & 1.00e-8 \\
F & 3.02e-8 & Cr & 4.68e-7 \\
Ne & 1.00e-4 & Mn & 2.88e-7 \\
Na & 2.14e-6 & Fe & 2.82e-5 \\
Mg & 3.47e-5 & Co & 8.32e-8 \\
Al & 2.95e-6 & Ni & 1.78e-6 \\
Si & 3.47e-5 & Cu & 1.62e-8 \\
P & 3.20e-7 & Zn & 3.98e-8
\enddata
\tablecomments{
{Solar abundances of different elements
relative to H, incorporated 
in $\Cloudy$--C17. Abundances 
of most elements are 
taken from \citet{1998SSRv...85..161G}, 
except for C, O which are taken 
from \citet{2002ApJ...573L.137A}, 
and N, Ne, Si, Mg, Fe are 
taken from \citet{2001AIPC..598...23H}.}}
\label{table:solar_abun}
\end{deluxetable}

\subsection{$\feii$ atomic datasets}
\begin{figure}[ht!]
\epsscale{1.2}
\plotone{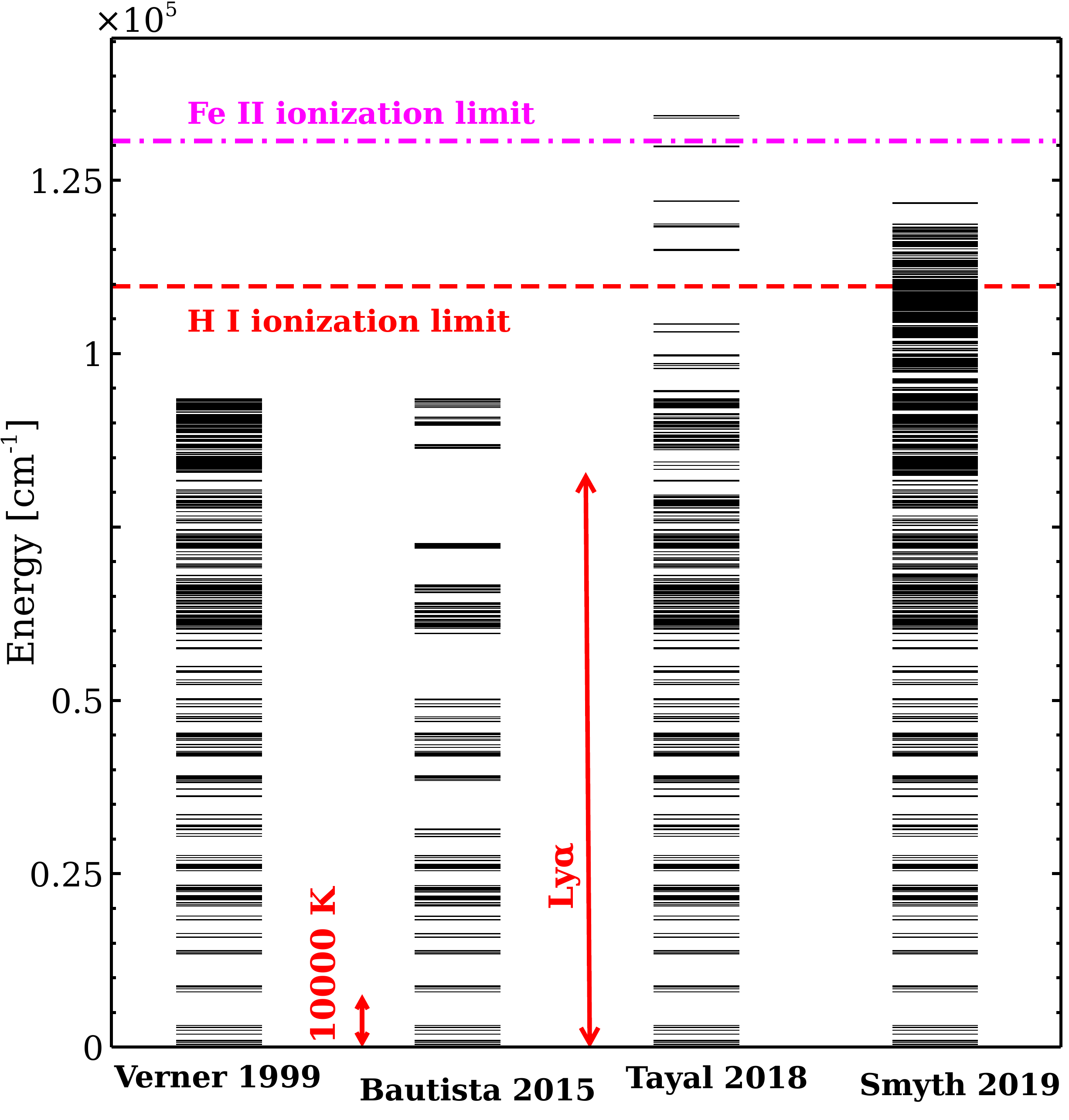}
\caption{Diagram showing the energy 
levels of various models of the $\feii$ atom. 
From left to right : 
dataset derived by \citet{1999ApJS..120..101V}, 
\citet{2015ApJ...808..174B}, 
\citet{2018PhRvA..98a2706T}, and 
\citet{2019MNRAS.483..654S}. 
The horizontal dashed lines indicate
the $\hi$ and $\feii$ ionization limits.
The energy of the 
Ly$\alpha$ transition, an important
source of photoexcitation of $\feii$, is also 
indicated,
along with the thermal energy corresponding to 10$^4$~K.
}
\label{fig:atomic}
\end{figure}

The $\feii$ ion has a complex
structure with 25 electrons, and 
is a ``grand challenge'' 
problem in atomic physics.
An accurate set of radiative 
and collisional atomic data is 
therefore needed 
to treat the selective excitation, 
continuum pumping, and 
fluorescence, which are known to 
be important for the $\feii$ emission 
\citep[e.g.,][]{2004ApJ...615..610B,Bruhweiler_2008,2012MNRAS.425..907J,2016ApJ...824..106W,2020MNRAS.494.1611N}.
Uncertainties in the atomic data 
have been a longstanding limitation 
in interpreting line intensities. 
Below we discuss four different 
$\feii$ atomic datasets 
that are now available in the $\Cloudy$,
while their energy levels are compared in  
Figure \ref{fig:atomic}.

First, we consider the widely used 
\citet{1999ApJS..120..101V} $\feii$ dataset
with 371 atomic levels producing 
13,157 emission lines with  
a highest energy level of 
$\sim$ 11.6 eV. 
\citet{1999ApJS..120..101V} 
data has collision strengths mainly 
calculated from the ``g-bar" approximation.
The original paper presented a 
total of $\sim 68,000$ transitions 
in their model atom. 
Most of these transitions   
are strongly forbidden and are 
assigned very small transition rates\footnote{ Beginning in C17 
\citep{2017RMxAA..53..385F} 
we only predict transitions which 
emit photons, accounting for 
the much smaller number of lines.}. 
Transition probabilities between 
their energy levels have uncertainties 
$\lesssim$ 20\% for strong permitted 
lines and $\gtsim$ 50\% for weak 
permitted and inter-combination lines. 
Also, forbidden lines have uncertainties
$\lesssim$ 50\%. We exclude all the totally 
forbidden lines 
from the $68,000$ transitions because 
of their small transition rates. 
Figure \ref{fig:atomic} shows the energy 
levels predicted by the 
\citet{1999ApJS..120..101V} $\feii$ model. 

\citet{2015ApJ...808..174B} reports 
a $\feii$ model with 159 levels 
extending up to 11.56 eV, 
producing 628 emission 
lines. 
Uncertainties in the transition 
probabilities lie between 10\% and 30\%. 
Tests show that the 
\citet{2015ApJ...808..174B} $\feii$
dataset has no lines between 
2000\AA\  and 3000\AA. 
As our paper mainly focuses on 
the UV band 
(2000\AA -- 3000\AA ) in $\feii$ 
spectra, 
we do not use this data-set for our 
BLR modelling. 

\citet{2018PhRvA..98a2706T} calculate 
340 energy levels with a highest 
energy
of $\sim$ 16.6 eV, including all 
levels from the $3d^64s$, $3d^54s^2$, 
$3d^7$, $3d^64p$
configurations and a few levels 
from the $3d^54s4p$ configuration. 
Transitions between these energy 
levels produce 57,635 emission lines 
with uncertainties in transition 
probabilities of
$\lesssim$ 30\% (in the 2200\AA--7800\AA). 
This dataset contains autoionizing 
levels with E $>$ 13.6 eV,  
which are absent in 
\citet{1999ApJS..120..101V} and 
\citet{2015ApJ...808..174B}. 
However, the density of states in 
high-lying energy levels are low, 
as shown in Figure \ref{fig:atomic}. 

Another larger $\feii$ data-set 
is also recently available in $\Cloudy$. 
\citet{2019MNRAS.483..654S} compute 
energy levels taking into account 
216 $LS$ terms in $\feii$ atom arising 
from the $3d^64s$, $3d^7$, 
$3d^64p$, $3d^54s^2$, and $3d^54s4p$ 
configurations. 
The \citet{2019MNRAS.483..654S} model 
has considerably more configurations (see pg 
657) in the structure but only includes 716 
levels in the close coupling (scattering 
model), with the highest energy level reaching
26.4 eV.
These levels produce 255,974 emission lines. 
The \citet{2019MNRAS.483..654S} 
dataset also contains autoionizing levels, 
but the density of states in the 
high-lying energy states is large compared 
to \citet{2018PhRvA..98a2706T}.

We model the $\feii$ emitting cloud using 
\citet{1999ApJS..120..101V}, 
\citet{2018PhRvA..98a2706T}, and 
\citet{2019MNRAS.483..654S} 
and compare their predictions. 
Results are discussed in Section \ref{sec:App}.

\subsection{AGN SEDs}
\begin{figure}[ht!]
\epsscale{1.2}
\hspace{-10pt}
\plotone{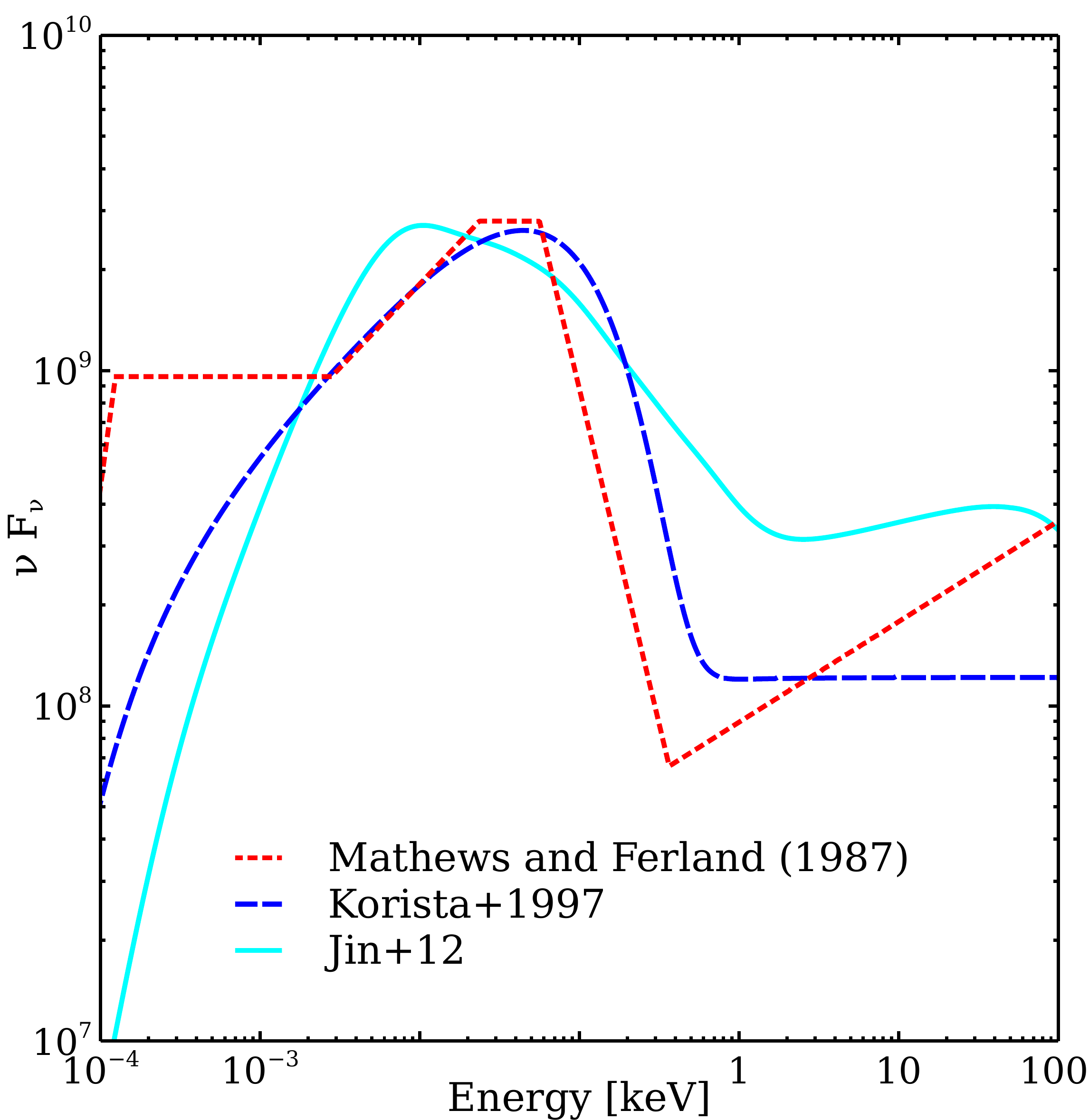}
\caption{Comparing different 
SEDs used in $\Cloudy$. 
Red dashed line: SED dervived by 
\citet{1987ApJ...323..456M}. 
Blue dashed line: SED derived by 
\citet{1997ApJS..108..401K}. Cyan solid line: 
\citet{2012MNRAS.425..907J} SED. }
\label{fig:SEDs}
\end{figure}

The BLR spectra in AGN mostly originate 
from the gas clouds photoionized by 
continuum radiation coming from an 
accretion disk around the central black hole. 
An accurate SED from the 
UV through to the soft X-ray band 
is therefore important to understand 
the $\feii$ emission 
\citep[e.g.,][]{1985ApJ...288...94W,1999ApJS..120..101V,1997ApJS..108..401K,1985ApJ...288...94W,2004ApJ...615..610B,Bruhweiler_2008,2012MNRAS.425..907J,2016ApJ...824..106W,2020MNRAS.494.1611N}. 
The soft X-ray part of the SED 
can penetrate into low ionization
regions to heat the gas and produce 
the $\feii$ emission by thermal collisions.
We consider the three SEDs presented by
\citet{1987ApJ...323..456M},
\citet{1997ApJS..108..401K}, and
\citet{2012MNRAS.425..907J}, as shown in 
Figure \ref{fig:SEDs}.
These have been normalised to have 
the same flux of ionizing photons, 
$\phi({\rm H}^0) = 10^{20}$ cm$^{-2}$ s$^{-1}$. 

\begin{figure*}
\gridline{\hspace{-20pt}\fig{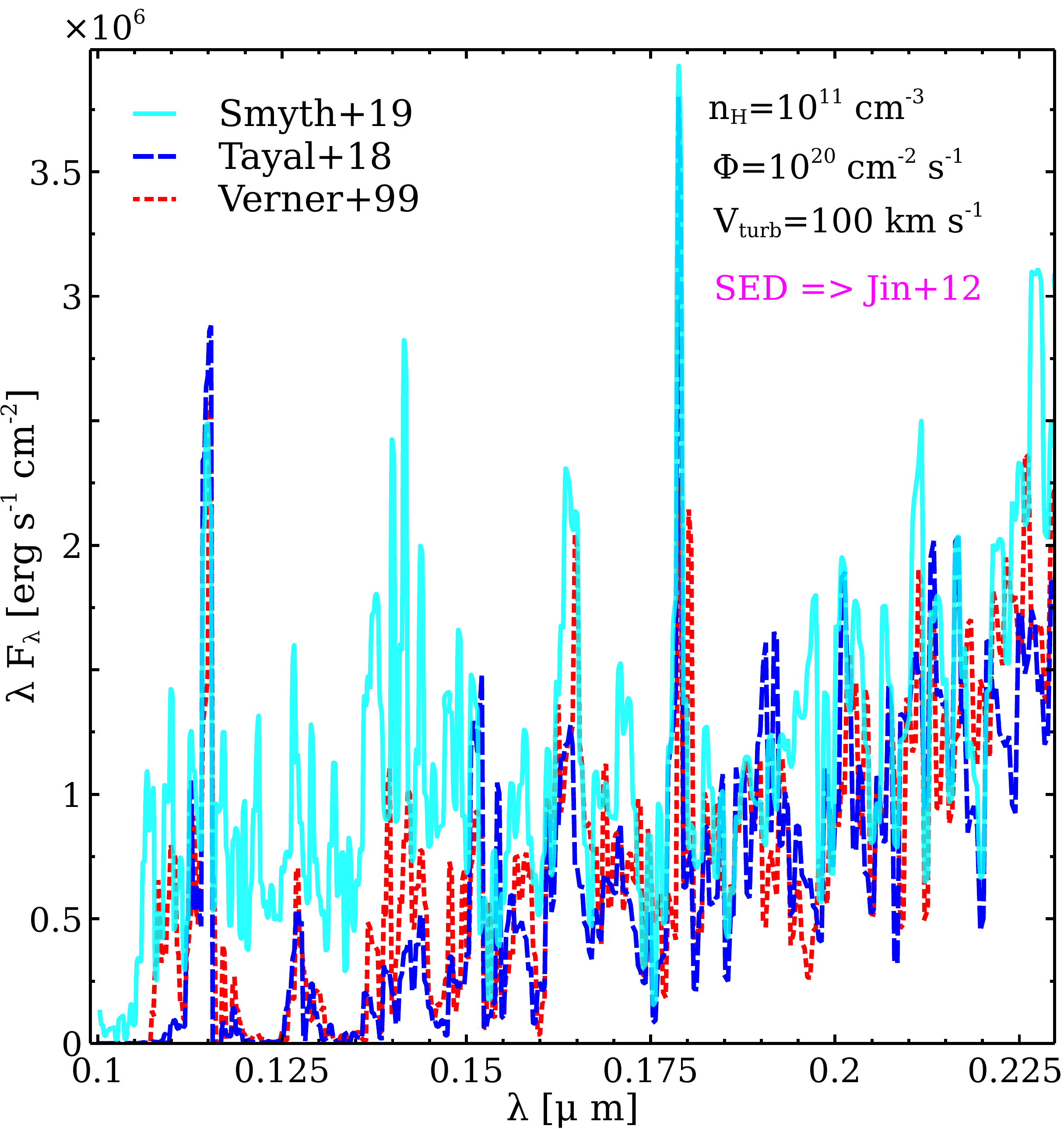}{0.33\textwidth}{(a)}
\hspace{-10pt}
          \fig{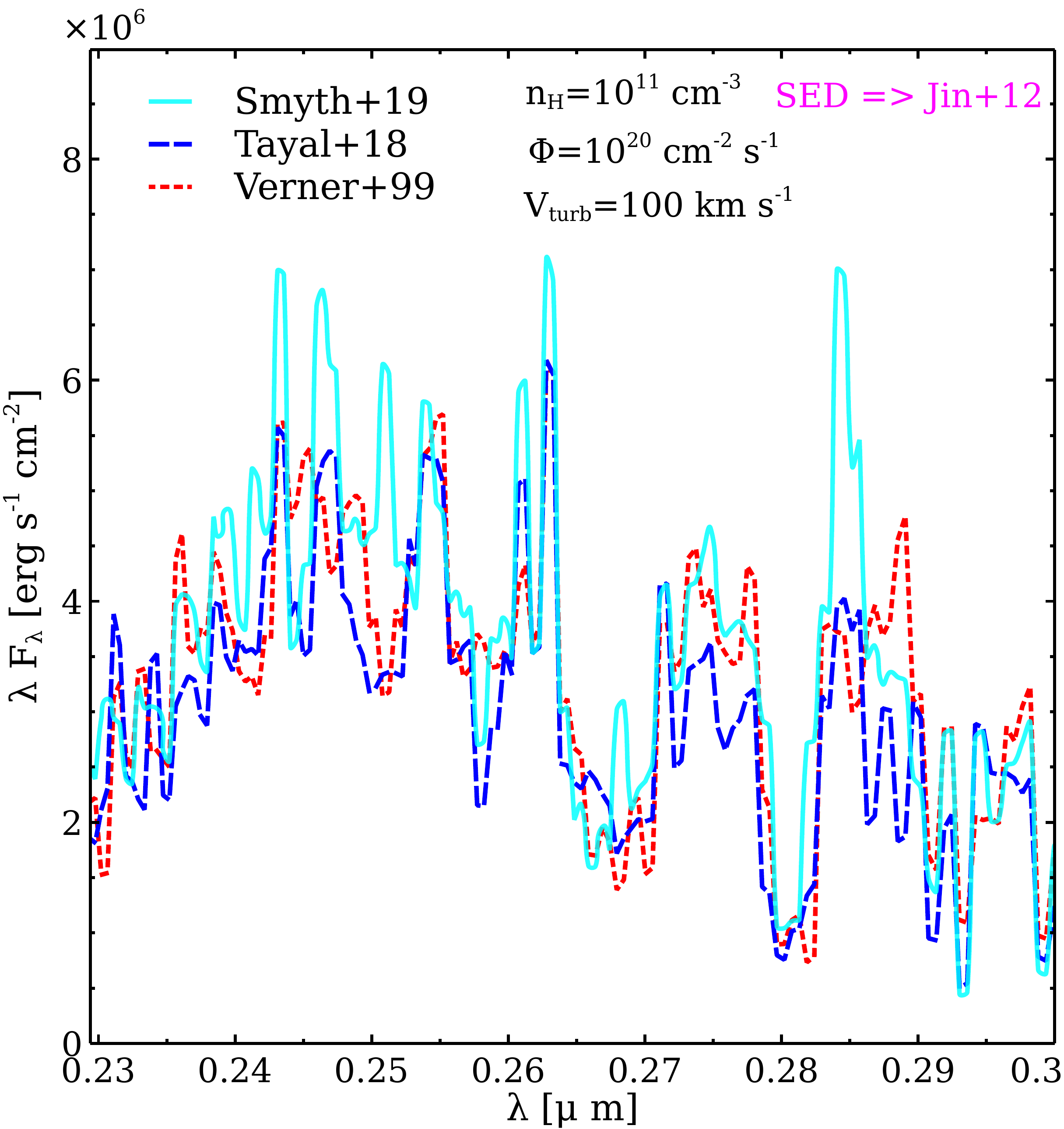}{0.33\textwidth}{(b)}
          \hspace{-10pt}
          \fig{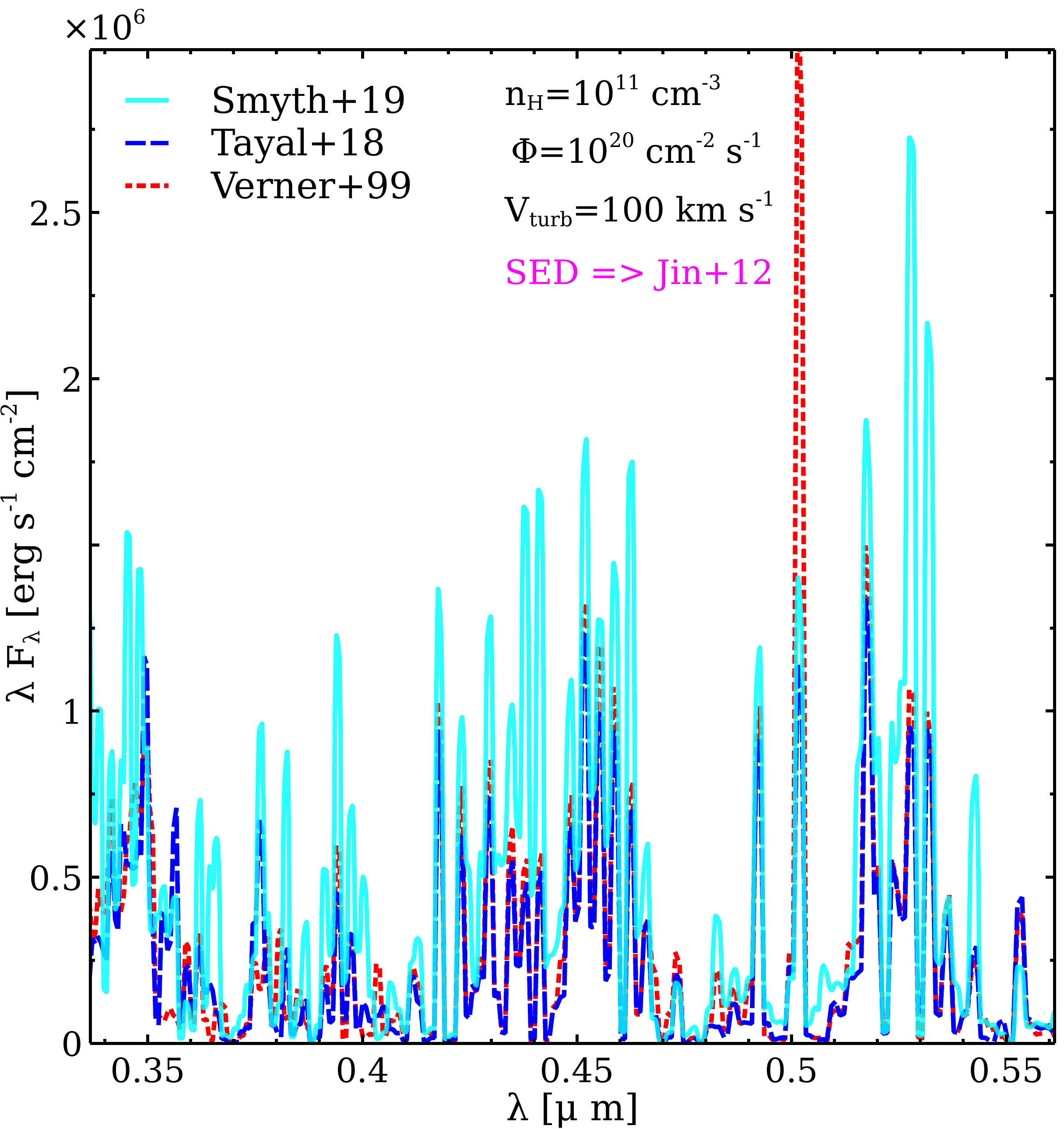}{0.33\textwidth}{(c)}}
\caption{Comparing $\Cloudy$ predicted 
$\feii$ spectra in 0.1--0.58 $\mu$m 
wavelength band 
using three different $\feii$ datasets 
and a fixed SED \citep{2012MNRAS.425..907J}. 
Red dotted line: $\feii$ spectra using 
\citet{1999ApJS..120..101V} dataset.
Blue dashed line: \citet{2018PhRvA..98a2706T} 
dataset. Cyan solid line: 
\citet{2019MNRAS.483..654S} dataset. 
Ranges 
of y-axis are different in all 
sub-figures.}
\label{fig:specs_dataset}
\end{figure*}

\citet{1987ApJ...323..456M}  derived 
a simple and phenomenological 
SED extending from the infrared 
(0.00124 eV) through the hard 
X-ray (10$^5$ eV), 
as shown in Figure \ref{fig:SEDs}. 
The shape of the continuum is 
approximated as a series of broken 
power-laws, i.e - 
\[{\rm f}_{\nu} = {\rm a}{\nu}^{-\alpha}, \]
where $\alpha$ is the spectral index. 
This can be determined by 
fitting the observed continuum with the 
power-law model in various energy bands. 

We consider another standard SED 
derived in \citet{1997ApJS..108..401K}, 
as shown in Figure \ref{fig:SEDs}. 
This is a baseline ionizing continuum 
model widely used to predict a wide range 
of emission lines in quasars 
\citep[e.g.,][]{2012JPhCS.372a2069R,2014AdSpR..54.1331M,2018MNRAS.478.5638M,2020MNRAS.496.2565T}. 
The derived shape of the continuum 
is a combination of a UV bump and an 
X-ray power law, i.e., 
\[{\rm f}_{\nu} = {\nu}^{-0.5}{\rm exp}(-{\rm h}{\nu}/{\rm kT}_{\rm {UV}}){\rm exp}(-{\rm kT}_{\rm IR}/{\rm h}{\nu})+{\rm a}{\nu}^{-1}, \]
where T$_{\rm UV}$ and T$_{\rm IR}$ are 
the 
cut-off temperatures of the UV bump 
and X-ray power law, respectively. 
The value of {`a'} can be determined 
from the ratio of the UV to X-ray 
continua, defined as- 
\[\frac{{\rm f}_{\nu}(2\ {\rm keV})}{{\rm f}_{\nu}(2500\A)} = 
403.3^{\alpha_{\rm ox}}, \]
where $\alpha_{\rm ox}$ distinguishes the 
continua between different type of AGNs. 
As an example, for Type I Seyfert 
galaxies, $\alpha_{\rm ox} = -1.2$. 
The shape of the continuum in the 
Figure \ref{fig:SEDs} corresponds 
to a UV bump peaking 
at $\sim$ 44 $\eV$ and an 
exponential decay part with 
a slope of $-2.3$.

We also consider the new 
generation of SEDs which use 
the Eddington
ratio (L/L$_{\rm Edd}$) as parameter and are
presented in \citet{2012MNRAS.425..907J} 
and summarized by 
\citet{2020MNRAS.494.5917F}.
These SEDs are a combination of theory and 
more recent observations.
{The SED with a log L/L$_{\rm Edd}$ = $-$0.55
is shown in Figure \ref{fig:SEDs}. }
The derived SED has three components, 
namely
(1) AGN disk emission, 
(2) Comptonization, and (3) a high 
energy 
power law tail \citep{2012MNRAS.420.1848D}. 
Figure \ref{fig:SEDs} represents 
the shape of the broadband SED. 
It increases as blackbody emission 
(emission from the outer disk) from lower 
energy, peaks in the UV, then falls off 
due to inverse Compton 
scattering in the inner disk and 
finally attains a power law tail 
which is due to the inverse 
Compton scattering in the corona.    
As shown, the SEDs are normalized 
to have the same total number of
ionizing photons.  
Note that 
they disagree
by $\sim 1$ dex in the soft 
X-ray region.


We next use all three SEDs in 
$\Cloudy$ and compare their 
predicted $\feii$ spectra.

\begin{figure*}
\gridline{\hspace{-20pt}\fig{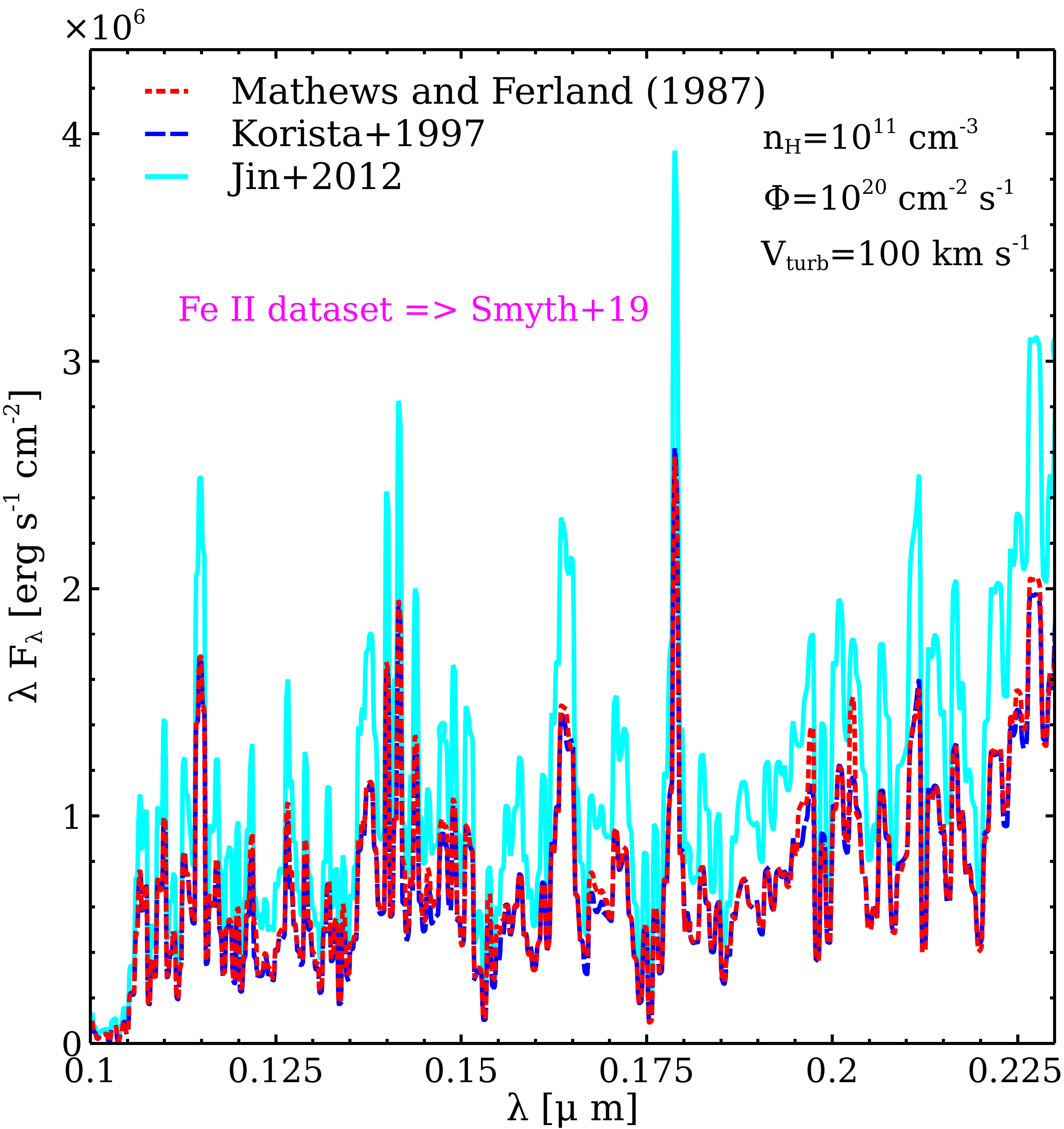}{0.33\textwidth}{(a)}
\hspace{-10pt}
          \fig{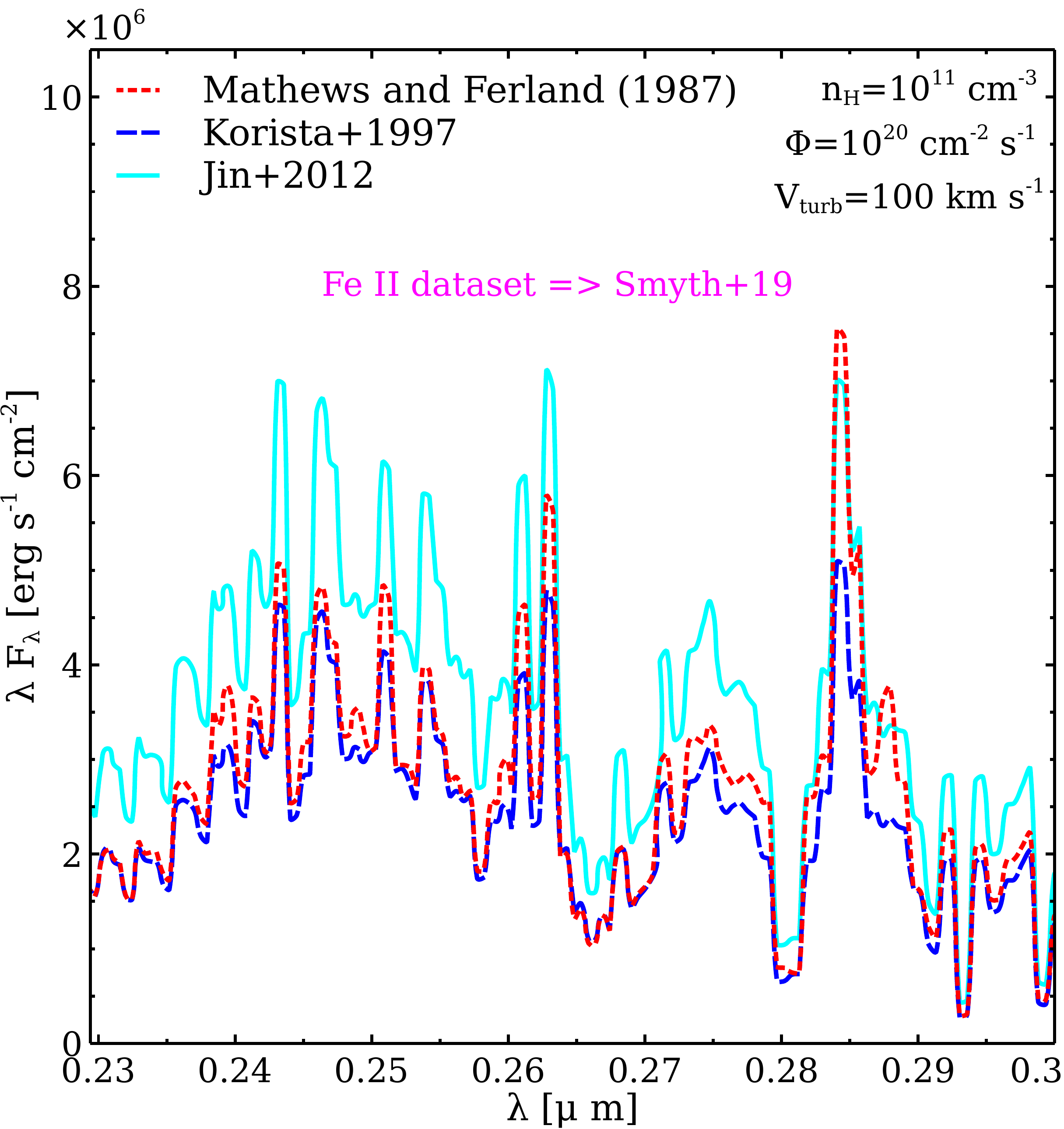}{0.33\textwidth}{(b)}
          \hspace{-10pt}
          \fig{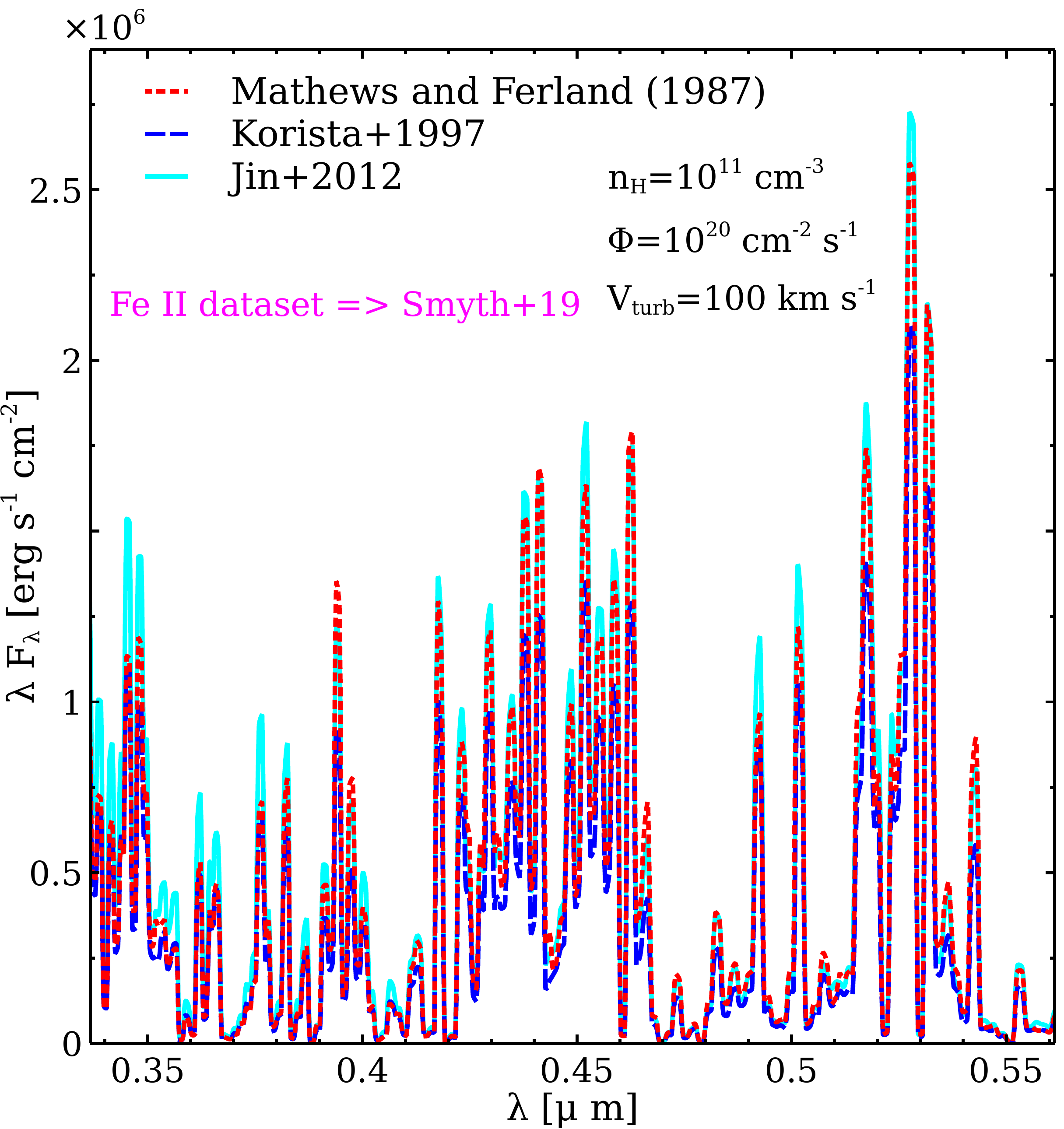}{0.33\textwidth}{(c)}}
\caption{Comparing $\Cloudy$ predicted 
$\feii$ spectra in 0.1--0.58 $\mu$m 
wavelength bands 
using three different AGN SEDs and a 
fixed $\feii$ dataset \citep{2019MNRAS.483..654S}. 
Red dotted line: $\feii$ spectra using 
\citet{1987ApJ...323..456M} SED.
Blue dashed line: \citet{1997ApJS..108..401K} SED. 
Cyan solid line: \citet{2012MNRAS.425..907J} SED. 
Ranges 
of y-axis are different in all sub-figures.}
\label{fig:spec_seds}
\end{figure*}
\begin{figure*}[ht!]
\epsscale{0.9}
\hspace{-10pt}
\plotone{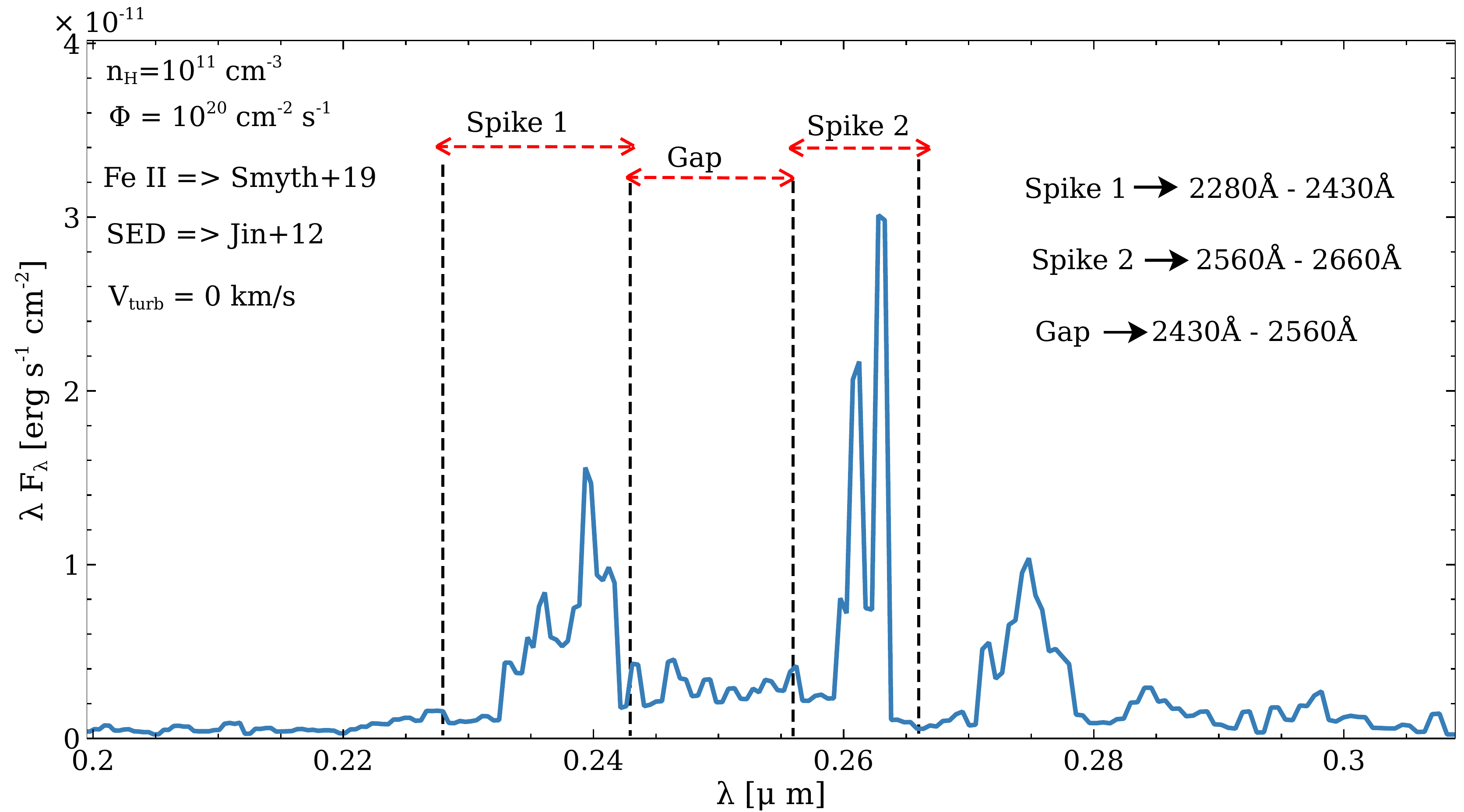}
\caption{Blue spectrum shows a typical 
$\feii$ spectra in UV using \citet{2019MNRAS.483..654S}
dataset, \citet{2012MNRAS.425..907J} SED. 
The Spikes and Gap are 
marked with vertical dashed lines.}
\label{fig:spike}
\end{figure*}

\section{Applications }\label{sec:App}

\subsection{Modelling of the BLR cloud}
We model the BLR gas by assuming solar 
abundances, as listed in table 
\ref{table:solar_abun}, and a 
cloud column density (N$_{\rm H}$) 
of $10^{24}$ cm$^{-2}$. 
First, we consider the $\feii$ emission 
from a single cloud modelled using 
the 
\citet{1999ApJS..120..101V}, 
\citet{2018PhRvA..98a2706T}, and 
\citet{2019MNRAS.483..654S} 
datasets in addition to an intermediate 
L/L$_{\rm Edd}$ AGN SED, as 
described in \citet{2012MNRAS.425..907J}. 
Figure \ref{fig:specs_dataset} 
compares the resulting $\feii$ emission 
spectra for the different atomic 
datasets in the UV and optical. 
In both spectral ranges, the 
\citet{2019MNRAS.483..654S} 
$\feii$ data-set produces larger 
line intensities compared 
to the other two. 
{
The \citet{2019MNRAS.483..654S} data-set has more
very highly excited states that connect by permitted 
transitions to 
low-excitation highly-populated levels.
Continuum fluorescent excitation is much stronger as a result,
which brings the short-wavelength lines into better agreement
with the template.} 
The \citet{2019MNRAS.483..654S} dataset
also better reproduces the observed $\feii$ 
pseudo-continuum, 
which results from the blending of a 
large number of high-lying  
lines in the UV and optical 
\citep{2012ApJ...751....7G}. 
We therefore 
select the \citet{2019MNRAS.483..654S} 
dataset to 
generate $\feii$ emission lines
throughout the remainder of this paper.

\begin{figure*}
\gridline{\hspace{-20pt}\fig{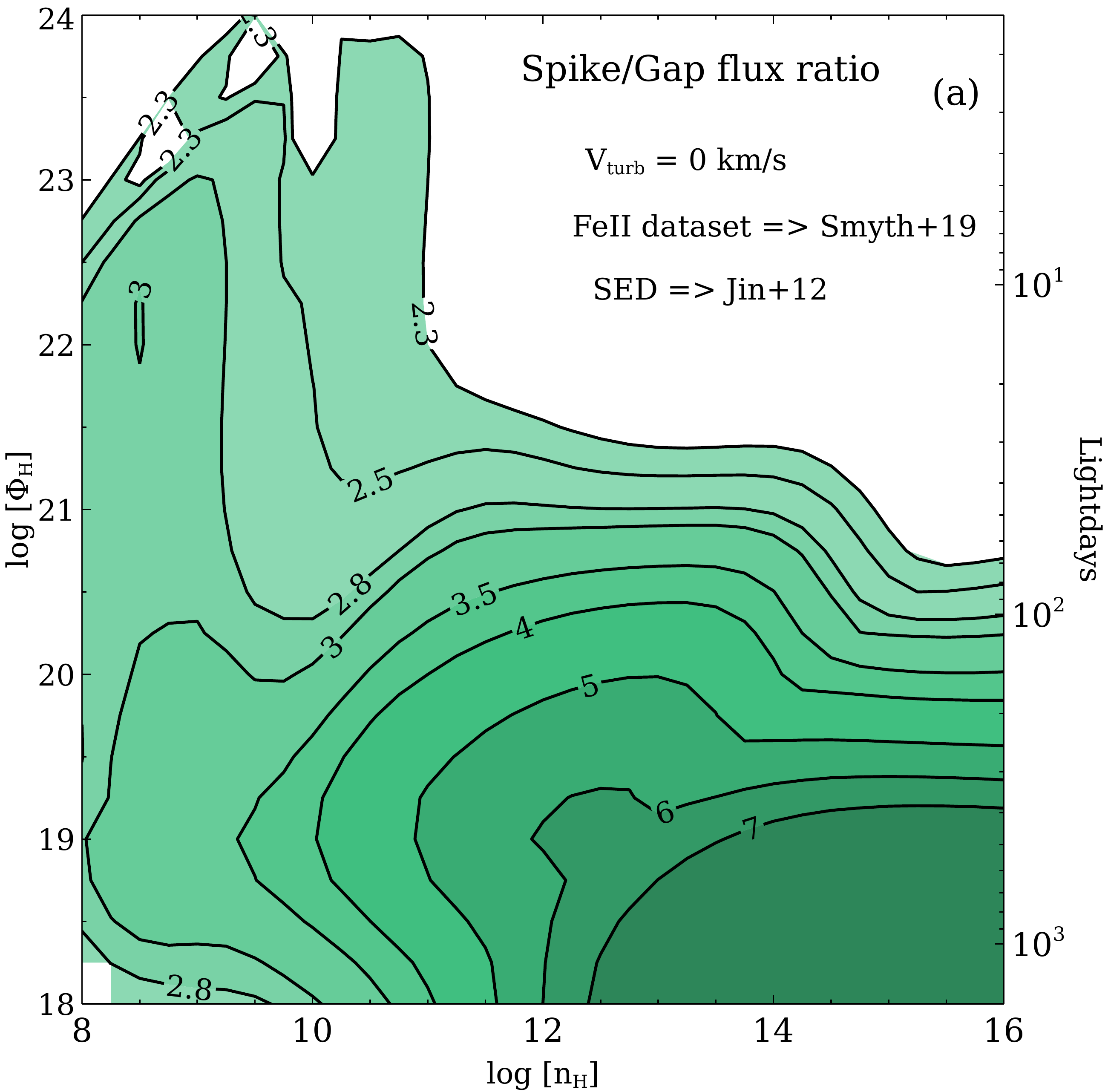}{0.54\textwidth}{}
\hspace{-1pt}
          \fig{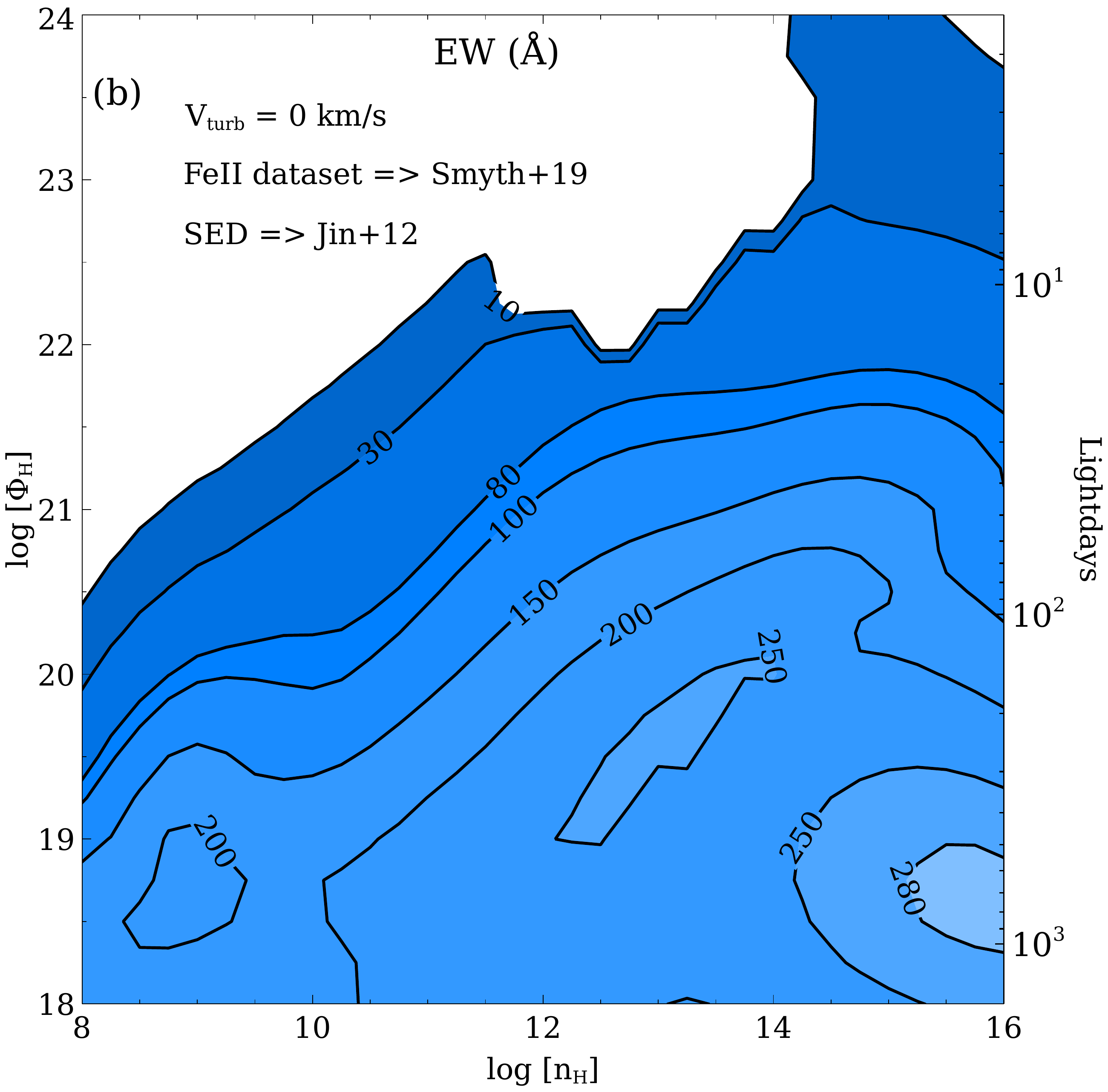}{0.54\textwidth}{}}
\caption{Contour plot of the Spike/gap 
ratio and equivalent width (EW) 
derived from 
photoionization model using the 
\citet{2019MNRAS.483..654S} $\feii$ 
dataset and 
\citet{2012MNRAS.425..907J} AGN SED. 
Left: Spike/gap ratio of the $\feii$ UV bump 
in the n$_{\rm H}$ -- $\Phi$ plane. 
Right: EW of $\feii$ 
UV bump in the n$_{\rm H}$ -- $\Phi$ plane.
An axis indicating the distance of $\feii$ 
emitting cloud from the central black hole 
is also added.}
\label{fig:contour_w/o_turb}
\end{figure*}

To find a suitable AGN SED for our 
BLR model, we predict $\feii$ spectra 
using
the \citet{1987ApJ...323..456M}, 
\citet{1997ApJS..108..401K}, and 
\citet{2012MNRAS.425..907J} SEDs in
addition to the \citet{2019MNRAS.483..654S} 
$\feii$ dataset.
Figure \ref{fig:spec_seds} compares 
the $\feii$ 
emission spectra for all three 
AGN SEDs we consider. 
The adopted density, flux, and 
turbulence parameters indicated in
the figures as are the best-fitting 
parameters derived below.

{The \citet{1987ApJ...323..456M} and
\citet{1997ApJS..108..401K} SEDs were simple empirical fits 
with little physical basis. In contrast,
\citet{2012MNRAS.425..907J} SED does have a theoretical 
foundation,
as explained in the original papers, and that family of SEDs 
produce
significantly more soft X-ray emission between 
$\sim$ 100 $\eV$ to 1 $\keV$ 
(see Figure \ref{fig:SEDs})
than the empirical SEDs.
Soft X-rays are important because these higher energy photons 
can 
penetrate further into the cloud, 
deposit more energy in neutral gas, 
and produce stronger $\feii$ 
lines by collisional excitation. 
Softer XUV and EUV\footnote{We refer to
the energy band
6 -- 13.6 eV (912\A\ -- 2000\A) as FUV,
13.6 -- 56.4 eV (228\A\ -- 912\A) as EUV,
and 56.4 -- few hundred eV ($<$ 228\A)
as XUV.} photons are extinguished at 
shallower depth into the cloud,
where Fe is more highly ionized, while hard X-rays and 
gamma-rays encounter
less opacity so are transmitted without being reprocessed into emission lines.}

{The two older SEDs are shown in 
Figure \ref{fig:spec_seds} for historical reference.
The papers describing them are 
highly cited (\citealt{1987ApJ...323..456M} and 
\citealt{1997ApJS..108..401K})
and they are built into $\Cloudy$.  
Several studies have used these SEDs, and the 
\citet{1999ApJS..120..101V} atomic data set,
to investigate  emission properties of AGN.
Although the older SEDs and the \citet{1999ApJS..120..101V}
data have historical 
interest, we prefer the modern SED \citep{2012MNRAS.425..907J}
with its
foundation of observations and theory and the new generation of 
atomic data \citep{2019MNRAS.483..654S} with 
more complete collision strengths.}

{For these reasons, we, therefore, adopt the \citet{2012MNRAS.425..907J} 
SED for our BLR model throughout this paper.
{Additionally,} the log L/L$_{\rm Edd}$ value 
for \citet{2012MNRAS.425..907J} 
SED is consistent with that of $\izw$ ($\gtrsim$ $-$0.60,  
\citealp[][]{2016MNRAS.462.1256C,2019A&A...630A..94G}).
This is the intermediate 
L/L$_{{\rm Edd}}$ SED and in 
Section \ref{sec:eign} we explore
the effects of varying L/L$_{{\rm Edd}}$.
We note that $\Cloudy$ makes it easy to adopt user-defined SEDs in new calculations.
It is hope that the work we present here will lead to further explorations
of the effects of the SED upon the line spectrum.
}


With the above selections of $\feii$ dataset 
and SED, 
the remaining parameters are the 
cloud hydrogen density (n$_{\rm H}$) 
and the 
flux of incoming photons striking the 
cloud ($\Phi$). 
We estimate these by considering a grid
of photoionization models by varying
these two parameters over a broad range. 

{
The computed $\feii$ spectra without microturbulence always show 
two prominent spikes at $\sim$ 2400\AA\ and 2600\AA, as 
extensively discussed by \citet{2004ApJ...615..610B} 
and shown in Figure \ref{fig:spike}.}
{These 
features are absent in the observed $\feii$ template.
Therefore, to compare the predicted 
$\feii$
spectra with the available observations of 
$\izw$ in the UV 
\citep{2001ApJS..134....1V}, 
we adopt the definitions of the Spike/gap ratio and 
$\feii$ equivalent width 
of the UV bump given in
\citet{2004ApJ...615..610B}. 
The ``Spike" is defined as the
total $\feii$ flux over the wavelength ranges 
2280\AA-2430\AA\ and 2560\AA-2660\AA\ (these are the 
two ``Spikes" as seen in Figure \ref{fig:spike}), 
and gap is defined as the total $\feii$ fluxes over 
2430\AA-2560\AA\ wavelength range.
The Spike/gap ratio is defined as the ratio of these two and
a typical observed 
value is $\approx$ 
1.4.}

We consider the equivalent width of 
the $\feii$ 
UV bump as the excess flux in 
2200\AA-2660\AA\ band over the 
continuum flux at 
1215\AA\ which is then 
divided by the continuum flux at 1215\AA. 
{$\Cloudy$ calculates the equivalent width
by assuming a 100\% covering factor\footnote{Covering 
factor or CF is defined as the fraction of 4$\pi$ sr 
covered by the clouds, as seen from the central 
black hole. Normally, the CF is expressed as 
$\Omega$/4$\pi$, where $\Omega$ is the solid angle.}. 
\citet{2004ApJ...615..610B} showed that a covering
factor of $\approx$ 20\% is a good indicator of a successful
model.
A typical $\Cloudy$ predicted
equivalent width of $>$ 400\AA\ is consistent 
with the observed value with a $\sim$ 20\% of covering
factor. }

Lines from the BLR are  velocity 
broadened (by $10^{3}-10^{4}$ km/s), although
$\izw$ has a line width of $\sim$ 1200 km/s. 
Given this, we 
smooth the $\Cloudy$ predicted spectra 
by a boxcar average 
with a width of 1200 km/s in order to 
compare with the observed \izw\ template.    


{We vary the hydrogen density 
(n$_{\rm H}$) from $10^8$ cm$^{-3}$ to $10^{16}$ 
cm$^{-3}$ and 
the photon flux ($\Phi$) 
between $10^{18}$ cm$^{-2}$ s$^{-1}$ 
and $10^{24}$ cm$^{-2}$ s$^{-1}$, 
as shown in Figure \ref{fig:contour_w/o_turb}, 
resulting in a total of} {825} {individual BLR 
models with a 10$^{0.25}$ step size.} 
{These parameters fully cover 
the possible range of clouds that produce the well-observed 
strong BLR lines like $\civ$ $\lambda$1549.}
In this first step we neglect any 
microturbulence inside the cloud
so the lines are thermally broadened. 
For each model, we analyze the 
$\feii$ emission spectrum and calculate 
the Spike/gap ratio and EW of 
the $\feii$ UV bump. 
Figure \ref{fig:contour_w/o_turb} shows  
contour plots of the Spike/gap ratio 
and the EW of the 
$\feii$ UV bump in the n$_{\rm H}-\Phi$ 
plane. 
No [n$_{\rm H}$, $\Phi$] pair reproduces the
observed $\feii$ spectra.
The \feii\ emission is too weak, and 
the Spike/gap ratio too large,
a problem \citet{2004ApJ...615..610B} 
also encountered.

We conclude that the standard 
baseline model cannot satisfactorily 
reproduce 
the observed UV $\feii$ emission even with 
a bigger $\feii$ atomic 
dataset and the new generation 
AGN SED. 
In addition to the Spike/gap ratio 
and equivalent width,
each baseline model also predicts 
two strong emission lines, 
one at 
2400\AA\  and another double-peaked 
at 2610\AA\  and 2630\AA , 
which are not present in the observed spectrum,
as shown in Figure \ref{fig:w_wo_turb}. 

\begin{figure}
\epsscale{1.4}
\plotone{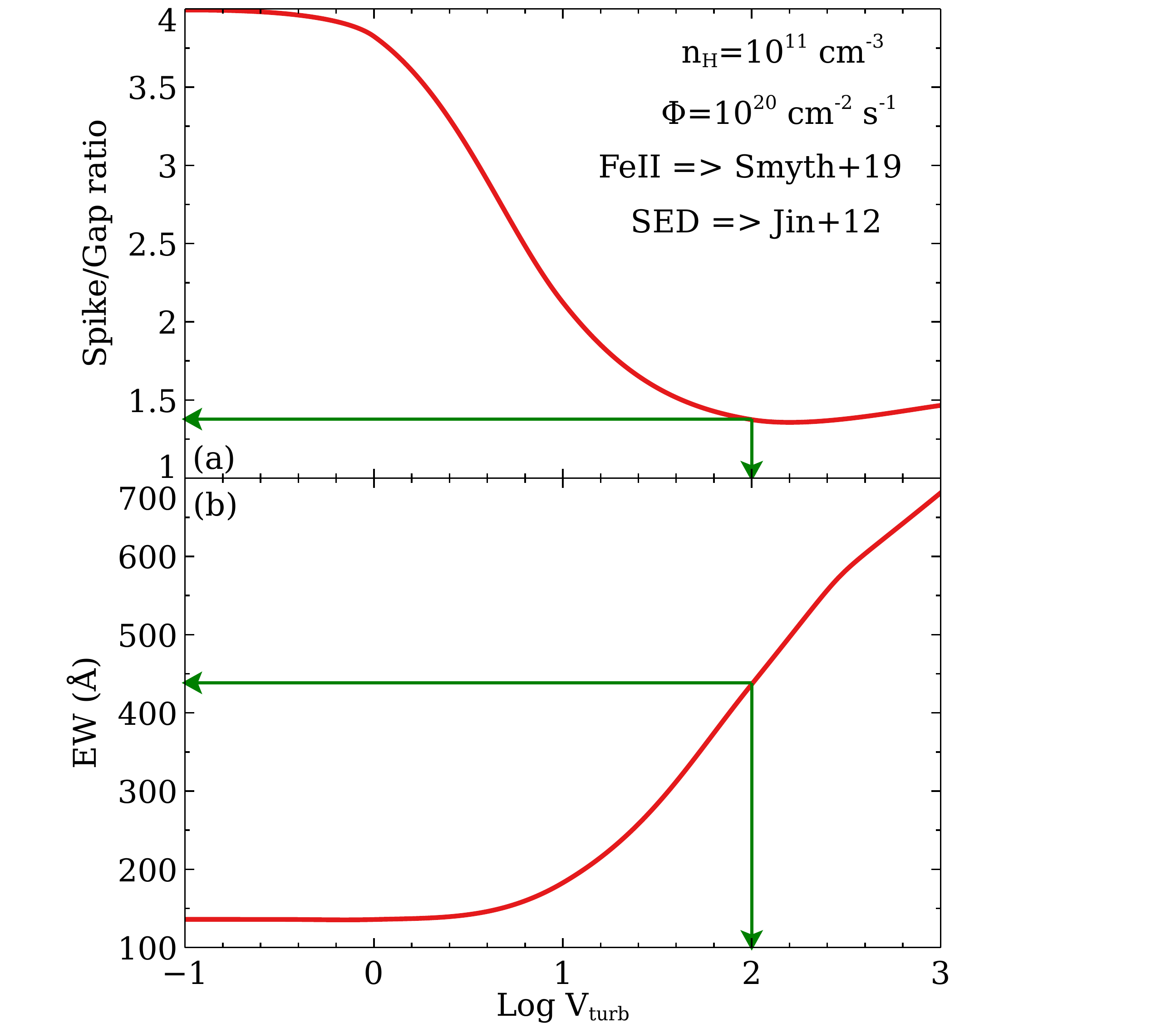}
\caption{Spike/gap ratio and EW 
of the $\feii$ emission in the UV as a 
function V$_{\rm turb}$. 
Green line: the Spike/gap 
ratio and EW of $\feii$ UV bump 
corresponding to V$_{\rm turb}$ = 100 km/s. 
These values are consistent with observation.
Any turbulence greater than 100 km/s 
will reproduce the observed \feii\ emission.
}
\label{fig:spike_EW}
\end{figure}

\subsection{The effects of microturbulence}
\begin{figure*}[ht!]
\gridline{\hspace{-20pt}\fig{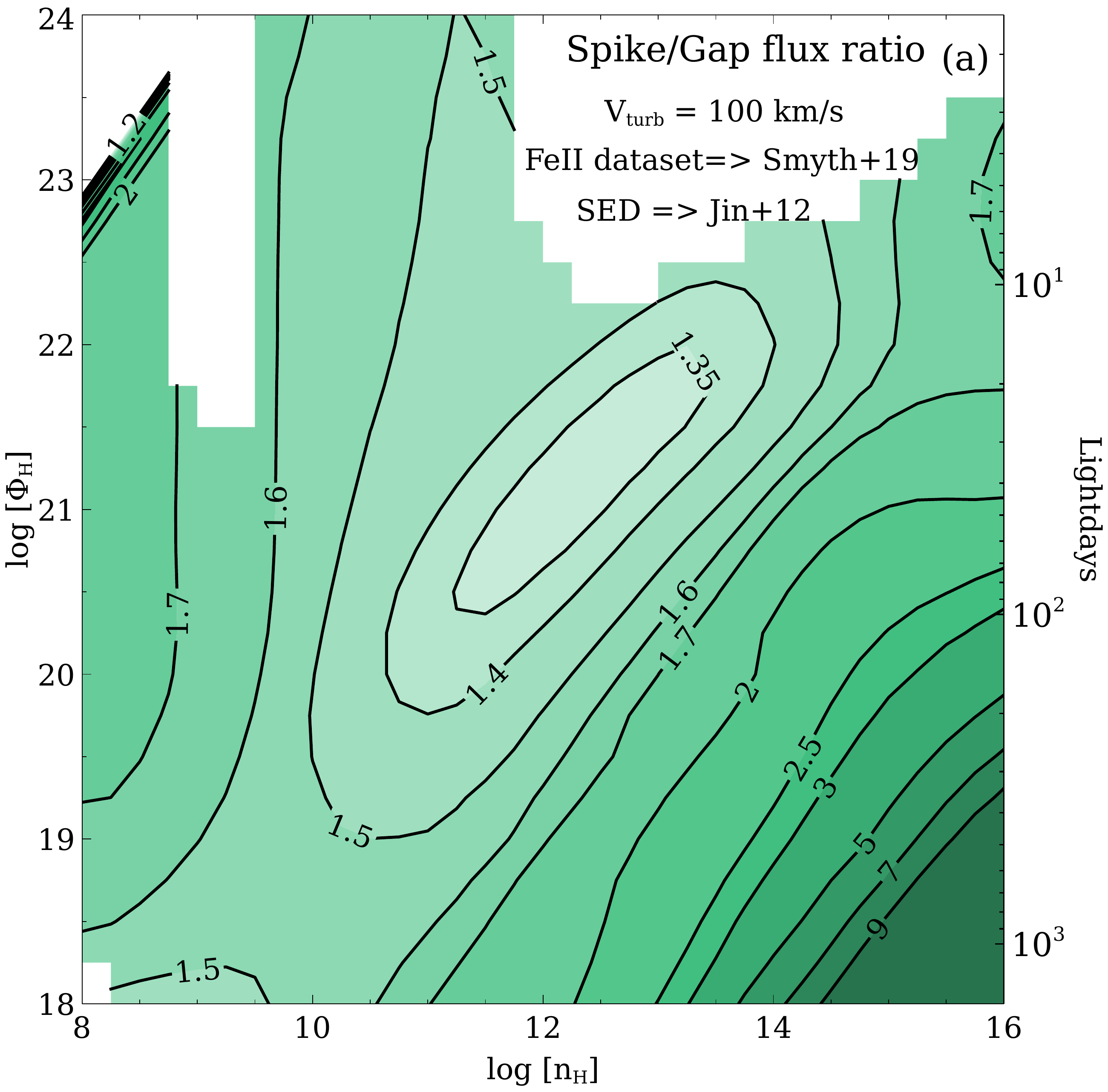}{0.54\textwidth}{}
\hspace{-1pt}
          \fig{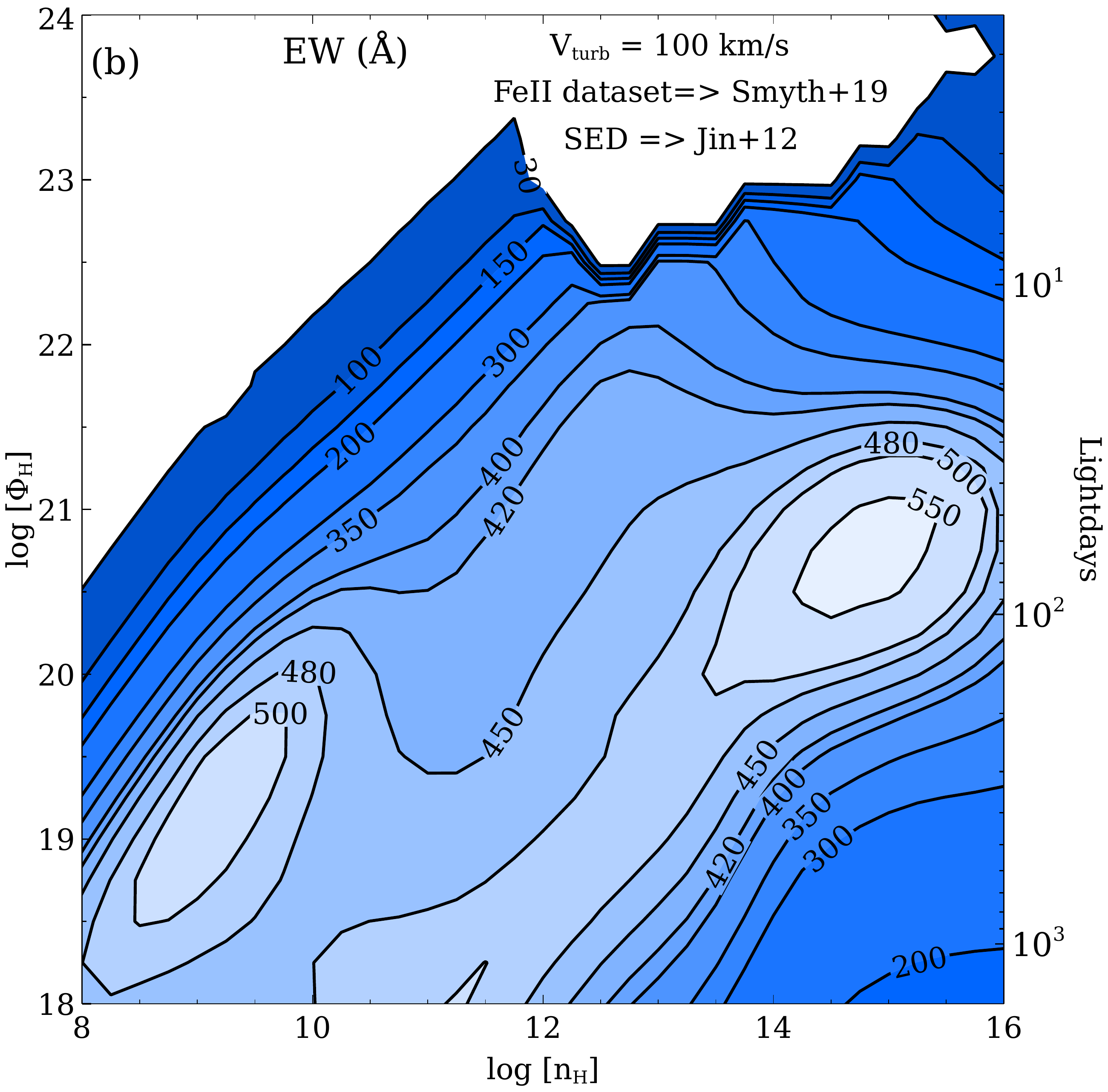}{0.54\textwidth}{}}
\caption{Similar to Figure 
\ref{fig:contour_w/o_turb} 
except V$_{\rm turb}$ = 100 km/s. 
Large regions of parameter space 
reproduce the observed Spike/Gap ratio
$\sim$ 1.4 and EW $>$ 400\AA\ of the 
\izw\ in the UV band.}
\label{fig:con_W_turb}
\end{figure*}

Next we consider clouds with non-zero 
microturbulent velocity
(V$_{\rm turb}$) throughout 
the BLR gas,
as previously proposed by 
\citet{1983ApJ...275..445N}, and later 
\citet{Bottorff_2000,2004ApJ...615..610B}. 
\citet{1983ApJ...275..445N}  showed that 
the strength of the $\feii$ UV emission is 
directly proportional to the microturbulence. 
The \citet{Bottorff_2000}, 
\citet{2004ApJ...615..610B}, and 
\citet{2008ApJ...675...83B} obtained 
a better fit for the quasar emission 
spectra, considering a turbulent emitting cloud. Physically, 
microturbulence decreases the optical 
depth by 
increasing the Doppler broadening
($\tau \sim V_{\rm doppler}^{-1}$), 
which helps more line photons to escape. 
Microturbulence also increases the 
importance of continuum pumping 
\citep[e.g.,][]{1977ApJ...215..733O,1978ApJS...38..187P, 1992ApJ...389L..63F}. 
We present a second grid model by 
varying V$_{\rm turb}$ between 
0.1 km/s and 10$^3$ km/s. 
The hydrogen density (n$_{\rm H}$) is 
fixed to 10$^{11}$ cm$^{-3}$ 
and the photon flux ($\Phi$) to 
10$^{20}$ cm$^{-2}$ s$^{-1}$, 
similar to the previous works 
\citep[e.g.,][]{2003ApJ...592L..59V,2004A&A...417..515V,2004ApJ...615..610B,Bruhweiler_2008,2020MNRAS.496.2565T}.
Figure \ref{fig:spike_EW} shows the 
Spike/gap ratio and the strength of 
$\feii$ emission in the UV as a function 
of V$_{\rm turb}$.
The Spike/gap ratio decreases with 
increasing V$_{\rm turb}$ and attains 
a value of $\sim$ 1.4 at 
$V_{\rm turb}\ > 60$ km/s.
In addition, to reproduce the observed 
strength of $\feii$ emission a 
V$_{\rm turb}$ $>$ 90 km/s is needed.
Hence we set V$_{\rm turb}$ = 100 km/s to 
reproduce the $\feii$ spectra for 
the remainder of the paper.

Next we recalculate the first grid of 
photoionization models assuming 
V$_{\rm turb}$ = 100 km/s 
to find the n$_{\rm H}$ and 
$\Phi$ of the emitting cloud. 
Similar to the previous steps, 
we calculate the Spike/gap 
ratio and the EW of the UV bump 
for each individual model. 
Figure \ref{fig:con_W_turb} 
presents  contour plots of the 
Spike/gap ratio and EW of the 
$\feii$ UV bump in the 
n$_{\rm H}$--$\Phi$ plane. 
Clearly, a cloud with n$_{\rm H}\ 
\approx\ 10^{11}$ cm$^{-3}$ and 
$\Phi\ \approx\ 10^{20}$ cm$^{-2}$ s$^{-1}$ 
reproduces the observed Spike/gap 
ratio and the strength of $\feii$ emission.

\begin{figure*}
\begin{tabular}{cc}
\includegraphics[width=0.7\textwidth]{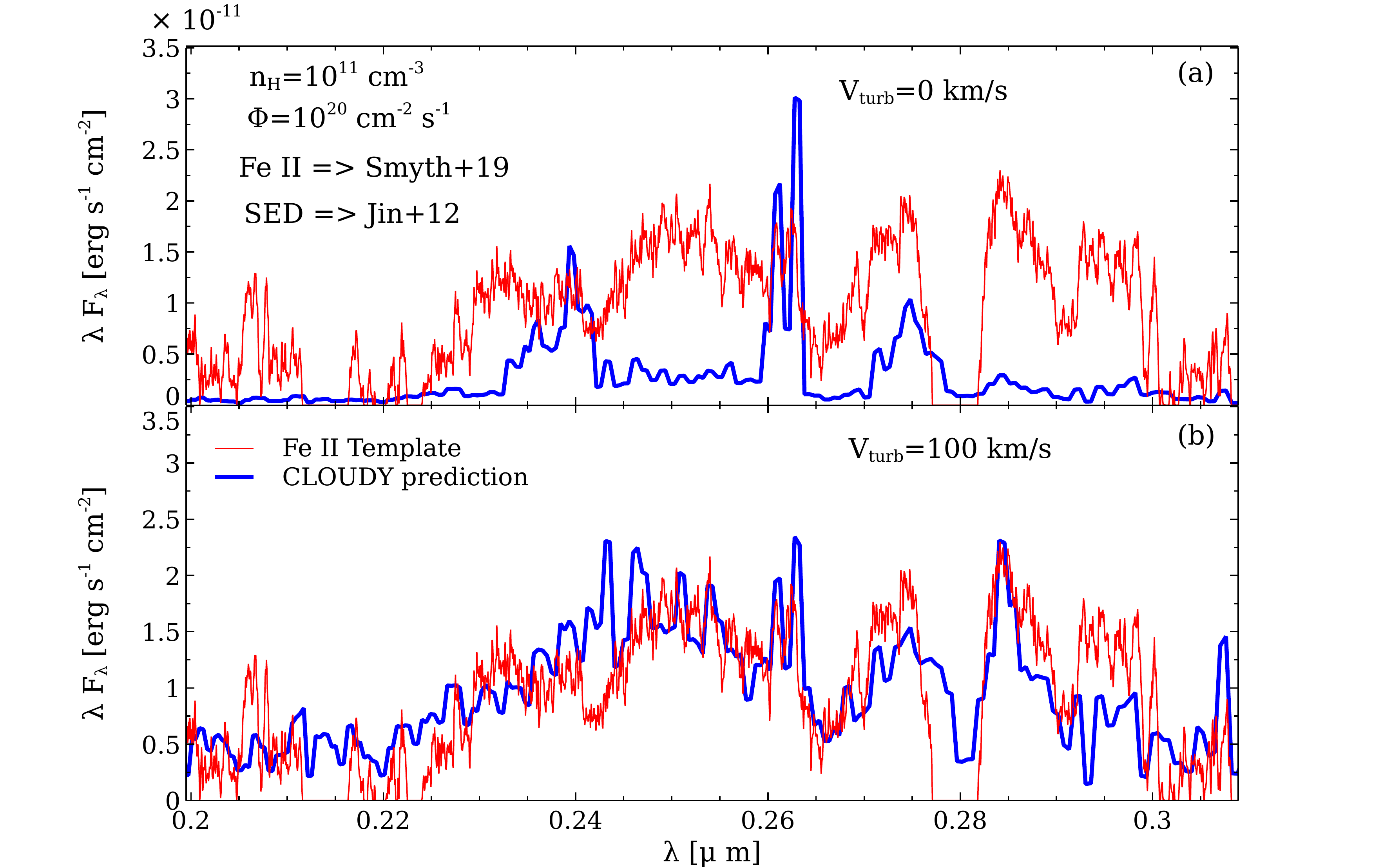} \\ \includegraphics[width=0.7\textwidth]{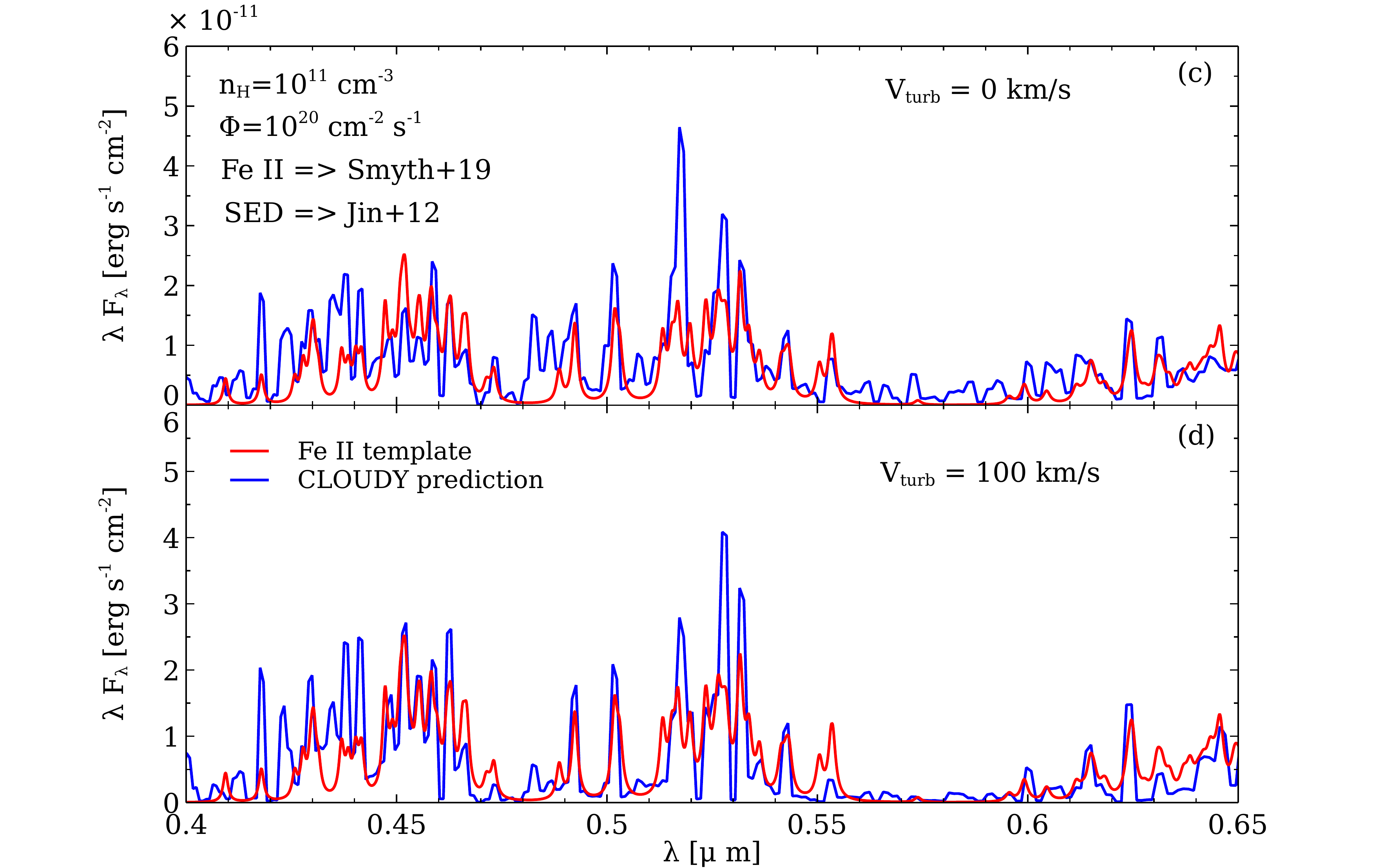}
\end{tabular}
\caption{{\sl Left}: panel (a) shows 
a comparison 
between the 
observed $\feii$ UV template 
\citep{2001ApJS..134....1V} 
and the $\Cloudy$ predicted $\feii$ UV spectrum
with 
V$_{\rm turb}$ = 0 km/s. Panel (b) shows the
same comparison with V$_{\rm turb}$ = 100 
km/s.
{\sl Right}: panel (c) compares the 
$\feii$ optical 
template \citep{2004A&A...417..515V} 
to the $\Cloudy$ 
predicted spectrum with 
V$_{\rm turb}$ = 0 km/s. 
Panel (d) presents 
the same comparison except 
V$_{\rm turb}$ = 100 
km/s. The turbulent model largely 
reproduces the template.}
\label{fig:w_wo_turb}
\end{figure*}

Next we compare the \feii\ emission 
in the UV and optical bands
with this density and flux.
In Figure \ref{fig:w_wo_turb} we compare the 
predicted 
spectra with the
observed $\feii$ templates in the UV 
\citep{2001ApJS..134....1V}
and optical \citep{2004A&A...417..515V}.
For further comparison, we include 
predictions for both thermal and 
turbulent clouds.
The \citet{2019MNRAS.483..654S}  dataset, 
the new-generation
SED \citep{2012MNRAS.425..907J}, 
and V$_{\rm turb}$ = 100 km/s 
reproduces AGN $\feii$ spectra in the UV and
optical far better than found in previous work, with 
solar Fe/H.
This is a remarkable accomplishment.

\subsection{$\feii$/$\mgii$ ratio}

The abundances of iron and magnesium are 
vital to probe the chemical evolution 
at high redshift. Iron in the solar 
neighborhood has mostly been produced by 
Type Ia SNe, which are 
the last stage of intermediate-mass 
stars in a close binary system. 
By comparison, magnesium has been 
deposited in the ISM by Type II SNe, 
a core-collapsed SN originating 
from the explosion of a massive 
star soon after the initial starburst. 
Type Ia SNe are delayed 
compared to Type II SNe. 
A time lag between the iron and magnesium 
enrichment of the solar neighborhood is 
therefore expected. This time 
delay varies from 0.3 Gyr for 
massive elliptical galaxies to 
1-3 Gyr for Milky-Way type galaxies 
\citep{2001ApJ...558..351M}. 
Previous observations have measured the 
$I(\feii)/I(\mgii)$ as a function of 
look-back time to trace the 
Fe/Mg abundance ratio 
\citep[e.g.,][]{2011ApJ...739...56D,2019ApJ...874...22S}.
{More recently \citet{2020arXiv201006902S} 
report on an extensive data set 
extending over a broad range of cosmic time
and find similar $I(\feii)/I(\mgii)$ ratios.}

We showed above that our best BLR model 
with solar abundances could successfully 
reproduce the observed template of \izw\  
in the UV and optical.
To estimate the $I(\feii)/I(\mgii)$, 
we consider the total flux 
of the $\feii$ UV bump 
between the 2000\AA--3000\AA\ range  
and that 
of the $\mgii$ doublet at 2798\AA\ . 
We examined the $I(\feii)/I(\mgii)$ for 
our best-fitting model with n$_{\rm H}$ = 
10$^{11}$ cm$^{-3}$
, $\Phi$ = 10$^{20}$ cm$^{-2}$ s$^{-1}$, and 
V$_{{\rm turb}}$ = 100 km/s.
Our model predicts
log($I(\feii)/I(\mgii)$) $\approx$ 0.7. 
\citet{2019ApJ...874...22S} obtained 
log($I(\feii)/I(\mgii)$) values of 0.31 -- 0.79
for a large sample of AGNs at $z\ \sim$ 3. 
\citet{2011ApJ...739...56D} showed a maximum 
log($I(\feii)/I(\mgii)$) of 0.77 for AGNs with 
$z\ >$ 4. Our log($I(\feii)/I(\mgii)$) 
predictions
closely match those observations, as shown in Figure \ref{fig:e}. 

\begin{figure}
\epsscale{1.2}
\plotone{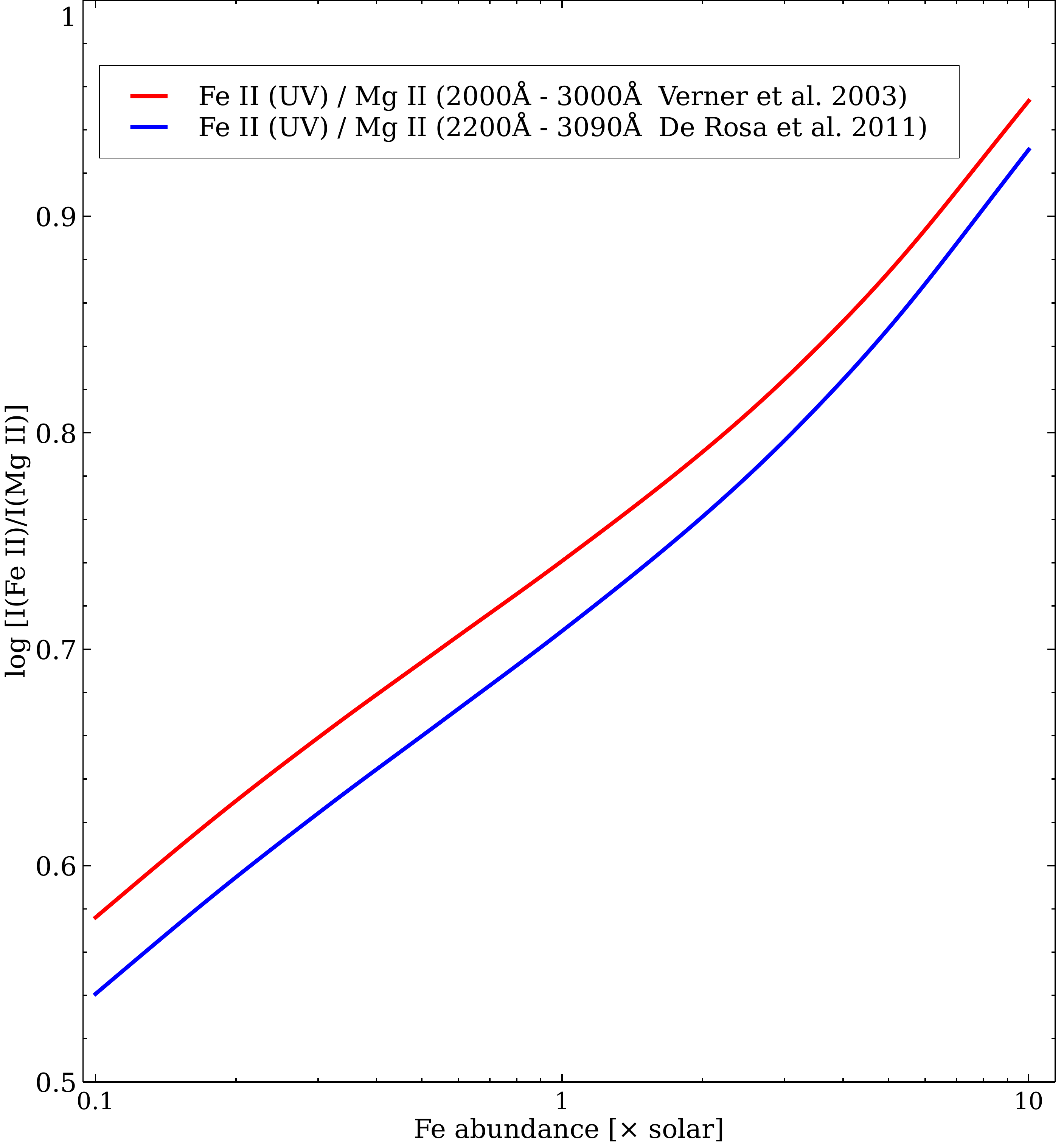}
\caption{$I(\feii)/I(\mgii)$ as 
a function of the Fe abundances. We show that
the $\feii$/$\mgii$ ratio for 
two different UV bands -- 
2000\AA\ - 3000\AA\ 
and 2200\AA\ - 3090\AA\ band, as described 
in \citet{2003ApJ...592L..59V} and 
\citet{2011ApJ...739...56D}, respectively.
}
\label{fig:abun_ratio}
\end{figure}

Next we vary the Fe abundance 
from 0.1 to 10 times the solar abundance 
in our 
best-fitting BLR model and estimate the 
$I(\feii)/I(\mgii)$ ratio while keeping the 
abundances of
the other elements constant.  This is a test of
the sensitivity of
the line intensity ratio to the abundance 
ratio.
Figure 
\ref{fig:abun_ratio} shows the 
monotonic increasing nature of the 
$I(\feii)/I(\mgii)$ ratio as a function 
of the Fe abundance. A similar trend of 
the $I(\feii)/I(\mgii)$ 
ratio with the Fe abundance 
is also obtained by 
\citet{2003ApJ...592L..59V}.
The log($I(\feii)/I(\mgii)$) intensity ratio 
increases by only $\sim 0.4$~dex when
the Fe/Mg abundance ratio increases by 2 dex,
or $I(\feii)/I(\mgii)$ $\propto$ 
(Fe/Mg)$^{0.19}$.
This shows that the \feii\ and \mgii\ spectra 
are strongly 
saturated due to the large
line optical depths so that the intensity ratio
does not strongly depend
on the abundance ratio.
Chemical evolution models 
suggest that the Fe/Mg ratio 
jumps by $\sim$ 1 dex when  
Type Ia supernovae start 
occurring in giant ellipticals
\citep{1999ARA&A..37..487H}. 
This would correspond to a change 
in the intensity ratio of log($I(\feii)/I(\mgii)$) $\sim$ 0.19.
Clearly a large number of high quality 
quasar spectra 
will be needed to 
measure such a subtle change.

\subsection{Eigenvector 1} \label{sec:eign}

\begin{figure*}
\begin{tabular}{cc}
\includegraphics[width=0.47\textwidth]{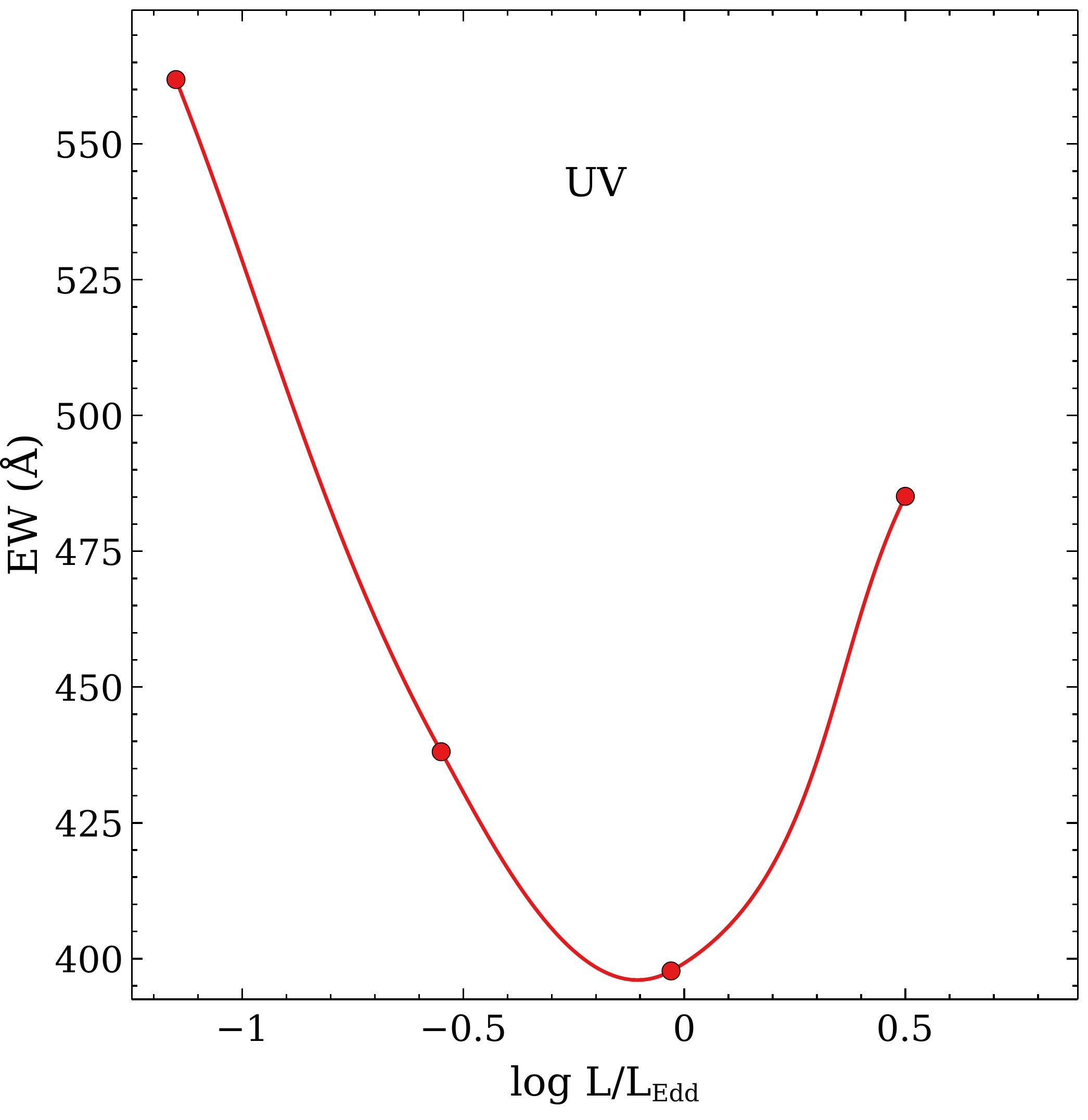}{(a)}   & \includegraphics[width=0.47\textwidth]{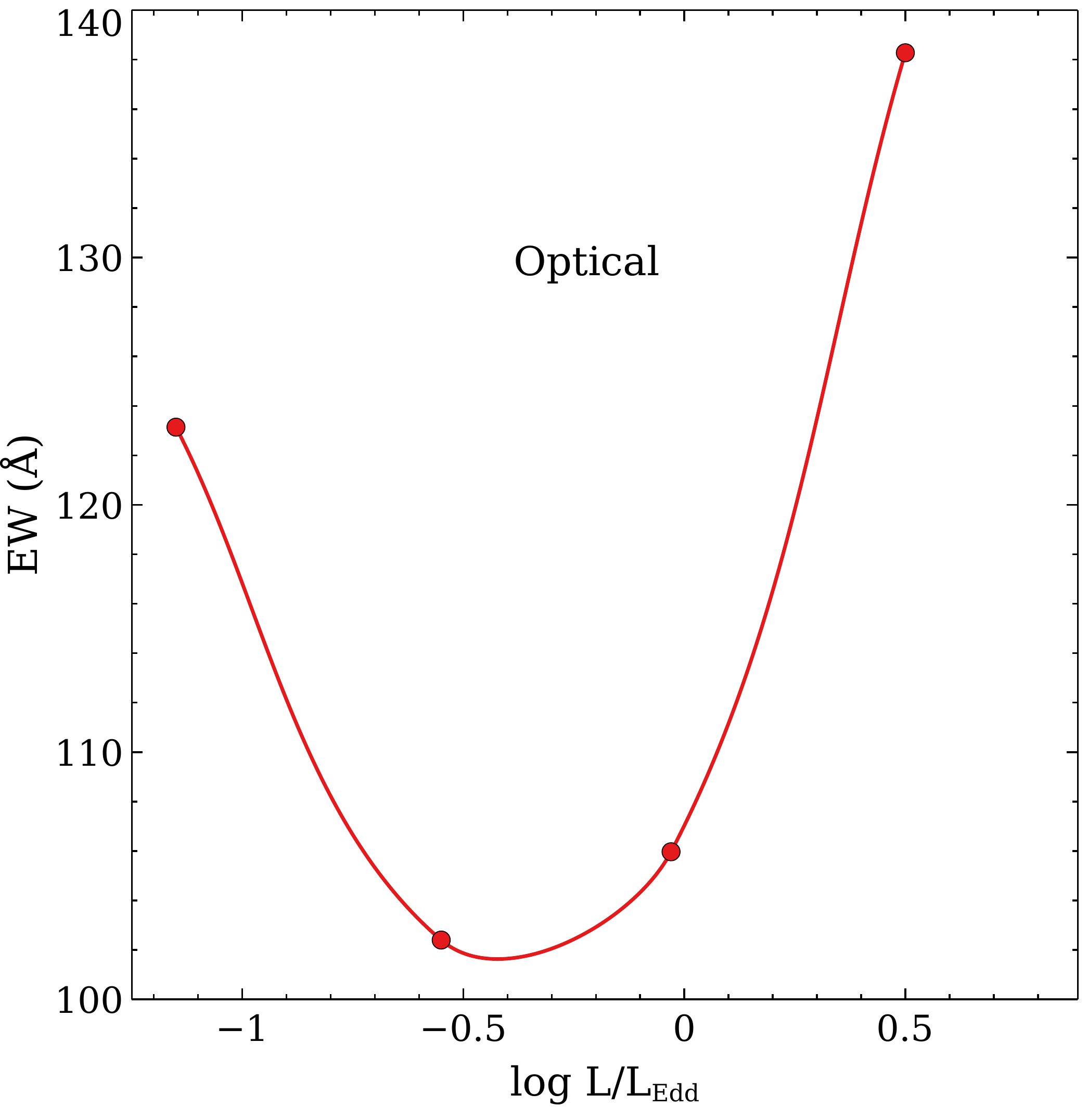}{(b)} \\
\includegraphics[width=0.47\textwidth]{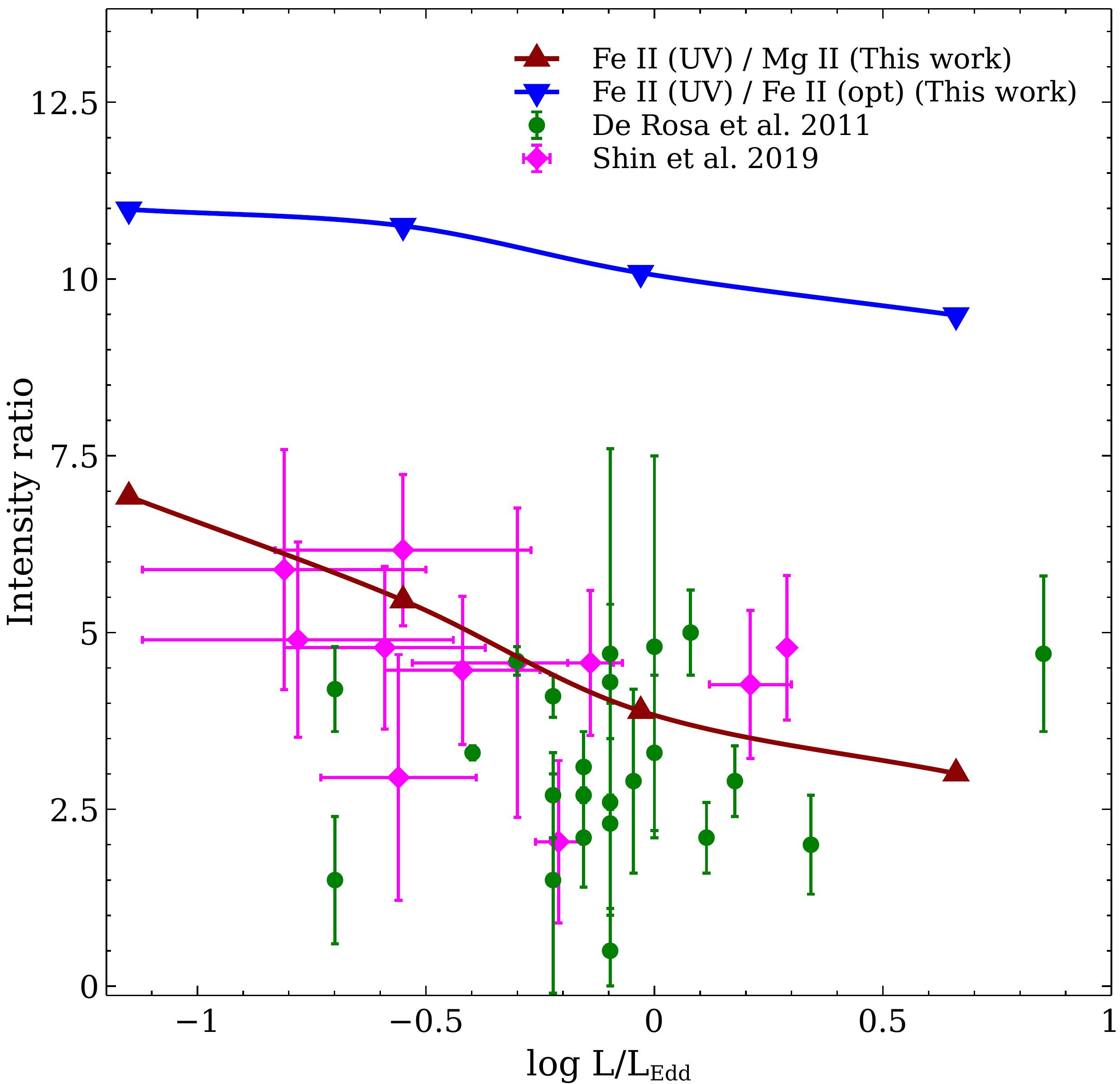}{(c)}\\
\end{tabular}
\caption{{Panel (a) and (b): red circles show
the EW of the $\feii$ UV bump 
and the $\feii$ optical band as a function of 
L/L$_{\rm Edd}$ respectively.
Panel (c): red and blue triangles 
represent the $I(\feii)/I(\mgii)$ and $I(\feii)(\rm 
UV)/I(\feii)(Optical)$
as a 
function of L/L$_{\rm Edd}$, respectively. 
For both plots
we use our best obtained model parameters.
Green circles and magenta squares are the 
observed $I(\feii)/I(\mgii)$ ratios taken from
\citet{2011ApJ...739...56D} and \citet{2019ApJ...874...22S} respectively.}}
\label{fig:e}
\end{figure*}

Next we consider the Eigenvector 1 correlation.
The new-generation SED models consider a 
wide range of L/L$_{{\rm Edd}}$,
as summarized by \citet{2020MNRAS.494.5917F}.
{We adopt four L/L$_{{\rm Edd}}$ values 
for \citet{2012MNRAS.425..907J} SED. Those are
log L/L$_{{\rm Edd}}$ = $-$1.15, $-$0.55, $-$0.03, and 0.66.}
Figure \ref{fig:e} shows the predicted 
change in the equivalent widths
of the UV bump and optical emission, 
defined as the total \feii\ emission
integrated over 4000 - 6000\AA.
Both are predicted as equivalent widths 
measured relative to the
incident SED at 1215\AA\ 
\citep{2004ApJ...615..610B}. 
The rightmost panel shows the 
$I(\feii)/I(\mgii)$ 
discussed above
and the ratio of the UV to optical \feii\ bands.  

{Although the predictions are in general agreement with observations,
the observational scatter makes comparison difficult}.
\citet{2011ApJ...739...56D} and 
\citet{2019ApJ...874...22S} obtained the
$I(\feii)/I(\mgii)$ for AGN samples
observed at $z\ >$ 4 and $z\ \sim$ 3, 
respectively. Their results show {little or}
no correlation between the $I(\feii)/I(\mgii)$ 
and L/L$_{{\rm Edd}}$.
Clearly there are more parameters changing 
than simply the SED.
A similar conclusion was reached by 
\citet{2020MNRAS.494.5917F}.
That study used analytical theory to 
predict how the equivalent widths
of $\hi$ and $\heii$ lines should change 
with the SED shape.
As expected, large increases in the 
equivalent width were predicted
as the SED grew harder with increasing 
L/L$_{{\rm Edd}}$.  
No changes in equivalent width are observed, 
showing that
the physics is more complex.
Alternatively, the underlying theory may not, 
for some reason, 
produce the correct change in the SED as 
L/L$_{{\rm Edd}}$ changes.
Clearly more work is needed.

\section{Conclusions and Summary}\label{summary}

We incorporated four atomic datasets, 
with self-consistent energy levels, 
transition rates, and 
electron impact collision strengths, 
for \feii.  
Three of these sets include
transitions in the UV and we compared
the emission 
line predictions from these datasets 
for typical 
AGN with a set of BLR parameters and 
the new generation of SEDs.
\begin{itemize}
    \item The original 
\citet{1999ApJS..120..101V} dataset 
largely rests on ``g-bar" 
collision strengths, so we only presented 
it as a reference.
By contrast, the \citet{2018PhRvA..98a2706T} 
and \citet{2019MNRAS.483..654S} $\feii$ 
datasets have many more 
transitions in the FUV, so 
continuum and Ly$\alpha$ florescent 
excitation, 
known to be important for \feii, 
is better reproduced. 
However, the \citet{2019MNRAS.483..654S} 
dataset, with its higher density 
of states, produces 
several times more emission 
in the FUV compared to      
\citet{2018PhRvA..98a2706T}. 
The  \citet{2019MNRAS.483..654S} 
dataset was hence adopted throughout most of 
this work.

\item We considered three representations of 
the broadband SED of an AGN, 
namely the original 
\citet{1987ApJ...323..456M} SED 
(long included in $\Cloudy$), 
the \citet{1997ApJS..108..401K} 
SED derived from observations, 
and that of \citet{2012MNRAS.425..907J} SED 
derived from a combination 
of theory and more recent observations. 
We compared the $\feii$ emission lines 
reproduced by each of the SEDs 
with $\Cloudy$. 
We showed that the \citet{2012MNRAS.425..907J} 
SED produces the 
strongest $\feii$ emission 
because the soft X-rays penetrate into the 
low ionization region, heating it,
generating more $\feii$ emission. 
{Also, the log L/L$_{\rm Edd}$ value for 
the intermediate \citet{2012MNRAS.425..907J} SED
is consistent with that of $\izw$.}
The \citet{2012MNRAS.425..907J} SED was 
hence adopted in this work.

\item We presented predictions of the strength of 
the broad bump of $\feii$ UV emission 
and the Spike/gap ratio, 
defined by \citet{2004ApJ...615..610B}, 
in the density (n$_{\rm H}$) -- photon flux ($\Phi$) plane, 
using solar abundances and thermal line 
broadening (V$_{\rm turb}$ = 0 km/s). 
Our calculations with the 
\citet{2019MNRAS.483..654S} 
dataset and the \citet{2012MNRAS.425..907J} 
SED also did not reproduce the observed 
$\feii$ template.

\item We considered the effects of 
microturbulence 
(V$_{\rm turb}$), 
varying it between 10$^{-1}$--10$^3$ km/s. 
We showed that a cloud with V$_{\rm turb}$ 
$>$
60 km/s closely 
reproduces the observed
Spike/gap ratio of 1.4, 
while the observed equivalent width ($>$ 400 \AA ) 
requires that V$_{\rm turb}$ $>$ 90 km/s. 
We adopted V$_{\rm turb}$ $\approx$ 100 km/s in 
this work.

\item Using the \citet{2019MNRAS.483..654S} dataset, 
\citet{2012MNRAS.425..907J} SED, solar 
abundances, 
and the derived V$_{\rm turb}$, 
we recalculated the strength of 
the $\feii$ UV emission and 
Spike/gap ratio in the density (n$_{\rm H}$) -- 
photon flux ($\Phi$) plane. 
We showed that large regions 
in the plane reproduced the observed UV  \feii\
emission. 


\item We compared our model using the 
best obtained BLR parameters 
with the observed $\feii$ UV 
\citep{2001ApJS..134....1V} 
and optical template 
\citet{2004A&A...417..515V} 
template and found 
surprisingly good agreement. 
We concluded that the $\feii$ 
spectra are consistent with formation 
in relatively dense 
(n$_{\rm H} = 10^{11}$ cm$^{-3}$) and  
turbulent (V$_{\rm turb}$ $\approx$ 100 km/s) 
clouds  
about  100 – 200 light days away from the 
central black hole with solar abundances. 

\item 
We showed that the 
$I(\feii)/I(\mgii)$ ratio predicted 
by our best-fitting model reproduces 
observations with solar abundances.
This is a significant step in calibrating 
\feii\ as an abundance indicator.
Unfortunately,  the
spectrum is strongly saturated so the intensity ratio
does not have a strong dependence on the abundance ratio.

\item We considered how changes in the SED 
shape, predicted
{from} recent models {by} varying L/L$_{\rm {Edd}}$, change
the resultant \feii\ spectrum.  
In the simplest case this would provide a physical explanation
for the observed Eigenvector 1 relations. 
The predicted changes do not agree with the observed correlations,
showing that more is changing than just the SED shape.  
A similar conclusion was reached by \citet{2020MNRAS.494.5917F}
in their analysis of hydrogen line equivalent widths.

\end{itemize}

\acknowledgments
GJF acknowledges support by NSF (1816537, 
1910687), 
NASA (ATP 17-ATP17-0141, 19-ATP19-0188), and 
STScI (HST-AR- 15018). 
MC acknowledges support by NSF (1910687), STScI
(HST-AR-14556.001-A), and NASA (19-ATP19-0188).
CAR, FPK and CPB 
are grateful to the UKRI Science and Technology
Facilities Research Council for support via 
grant ST/P000312/1.
{Finally, we thank the anonymous referee for his or her
helpful suggestions.}

\bibliography{sample63}{}

\begin{thebibliography}{}
\expandafter\ifx\csname natexlab\endcsname\relax\def\natexlab#1{#1}\fi
\providecommand{\url}[1]{\href{#1}{#1}}
\providecommand{\dodoi}[1]{doi:~\href{http://doi.org/#1}{\nolinkurl{#1}}}
\providecommand{\doeprint}[1]{\href{http://ascl.net/#1}{\nolinkurl{http://ascl.net/#1}}}
\providecommand{\doarXiv}[1]{\href{https://arxiv.org/abs/#1}{\nolinkurl{https://arxiv.org/abs/#1}}}

\bibitem[{{Allende Prieto} {et~al.}(2002){Allende Prieto}, {Lambert}, \&
  {Asplund}}]{2002ApJ...573L.137A}
{Allende Prieto}, C., {Lambert}, D.~L., \& {Asplund}, M. 2002, \apjl, 573,
  L137, \dodoi{10.1086/342095}

\bibitem[{{Baldwin} {et~al.}(1995){Baldwin}, {Ferland}, {Korista}, \&
  {Verner}}]{1995ApJ...455L.119B}
{Baldwin}, J., {Ferland}, G., {Korista}, K., \& {Verner}, D. 1995, \apjl, 455,
  L119, \dodoi{10.1086/309827}

\bibitem[{{Baldwin} {et~al.}(2004){Baldwin}, {Ferland}, {Korista}, {Hamann}, \&
  {LaCluyz{\'e}}}]{2004ApJ...615..610B}
{Baldwin}, J.~A., {Ferland}, G.~J., {Korista}, K.~T., {Hamann}, F., \&
  {LaCluyz{\'e}}, A. 2004, \apj, 615, 610, \dodoi{10.1086/424683}

\bibitem[{{Bautista} {et~al.}(2015){Bautista}, {Fivet}, {Ballance}, {Quinet},
  {Ferland}, {Mendoza}, \& {Kallman}}]{2015ApJ...808..174B}
{Bautista}, M.~A., {Fivet}, V., {Ballance}, C., {et~al.} 2015, \apj, 808, 174,
  \dodoi{10.1088/0004-637X/808/2/174}

\bibitem[{{Bolton} \& {Haehnelt}(2007)}]{2007MNRAS.374..493B}
{Bolton}, J.~S., \& {Haehnelt}, M.~G. 2007, \mnras, 374, 493,
  \dodoi{10.1111/j.1365-2966.2006.11176.x}

\bibitem[{{Boroson} \& {Green}(1992)}]{1992ApJS...80..109B}
{Boroson}, T.~A., \& {Green}, R.~F. 1992, \apjs, 80, 109,
  \dodoi{10.1086/191661}

\bibitem[{Bottorff {et~al.}(2000)Bottorff, Ferland, Baldwin, \&
  Korista}]{Bottorff_2000}
Bottorff, M., Ferland, G., Baldwin, J., \& Korista, K. 2000, The Astrophysical
  Journal, 542, 644, \dodoi{10.1086/317051}

\bibitem[{{Bruhweiler} \& {Verner}(2008)}]{2008ApJ...675...83B}
{Bruhweiler}, F., \& {Verner}, E. 2008, \apj, 675, 83, \dodoi{10.1086/525557}

\bibitem[{Bruhweiler \& Verner(2008)}]{Bruhweiler_2008}
Bruhweiler, F., \& Verner, E. 2008, The Astrophysical Journal, 675, 83,
  \dodoi{10.1086/525557}

\bibitem[{{Cracco} {et~al.}(2016){Cracco}, {Ciroi}, {Berton}, {Di Mille},
  {Foschini}, {La Mura}, \& {Rafanelli}}]{2016MNRAS.462.1256C}
{Cracco}, V., {Ciroi}, S., {Berton}, M., {et~al.} 2016, \mnras, 462, 1256,
  \dodoi{10.1093/mnras/stw1689}

\bibitem[{{De Rosa} {et~al.}(2011){De Rosa}, {Decarli}, {Walter}, {Fan},
  {Jiang}, {Kurk}, {Pasquali}, \& {Rix}}]{2011ApJ...739...56D}
{De Rosa}, G., {Decarli}, R., {Walter}, F., {et~al.} 2011, \apj, 739, 56,
  \dodoi{10.1088/0004-637X/739/2/56}

\bibitem[{{Djorgovski} {et~al.}(2001){Djorgovski}, {Castro}, {Stern}, \&
  {Mahabal}}]{2001ApJ...560L...5D}
{Djorgovski}, S.~G., {Castro}, S., {Stern}, D., \& {Mahabal}, A.~A. 2001,
  \apjl, 560, L5, \dodoi{10.1086/324175}

\bibitem[{{Done} {et~al.}(2012){Done}, {Davis}, {Jin}, {Blaes}, \&
  {Ward}}]{2012MNRAS.420.1848D}
{Done}, C., {Davis}, S.~W., {Jin}, C., {Blaes}, O., \& {Ward}, M. 2012, \mnras,
  420, 1848, \dodoi{10.1111/j.1365-2966.2011.19779.x}

\bibitem[{{Fan} {et~al.}(2006){Fan}, {Carilli}, \&
  {Keating}}]{2006ARA&A..44..415F}
{Fan}, X., {Carilli}, C.~L., \& {Keating}, B. 2006, \araa, 44, 415,
  \dodoi{10.1146/annurev.astro.44.051905.092514}

\bibitem[{{Ferland}(1992)}]{1992ApJ...389L..63F}
{Ferland}, G.~J. 1992, \apjl, 389, L63, \dodoi{10.1086/186349}

\bibitem[{{Ferland} {et~al.}(2020){Ferland}, {Done}, {Jin}, {Landt}, \&
  {Ward}}]{2020MNRAS.494.5917F}
{Ferland}, G.~J., {Done}, C., {Jin}, C., {Landt}, H., \& {Ward}, M.~J. 2020,
  \mnras, 494, 5917, \dodoi{10.1093/mnras/staa1207}

\bibitem[{{Ferland} {et~al.}(2009){Ferland}, {Hu}, {Wang}, {Baldwin}, {Porter},
  {van Hoof}, \& {Williams}}]{2009ApJ...707L..82F}
{Ferland}, G.~J., {Hu}, C., {Wang}, J.-M., {et~al.} 2009, \apjl, 707, L82,
  \dodoi{10.1088/0004-637X/707/1/L82}

\bibitem[{{Ferland} {et~al.}(2017){Ferland}, {Chatzikos}, {Guzm{\'a}n},
  {Lykins}, {van Hoof}, {Williams}, {Abel}, {Badnell}, {Keenan}, {Porter}, \&
  {Stancil}}]{2017RMxAA..53..385F}
{Ferland}, G.~J., {Chatzikos}, M., {Guzm{\'a}n}, F., {et~al.} 2017, \rmxaa, 53,
  385.
\newblock \doarXiv{1705.10877}

\bibitem[{{Garcia-Rissmann} {et~al.}(2012){Garcia-Rissmann},
  {Rodr{\'\i}guez-Ardila}, {Sigut}, \& {Pradhan}}]{2012ApJ...751....7G}
{Garcia-Rissmann}, A., {Rodr{\'\i}guez-Ardila}, A., {Sigut}, T.~A.~A., \&
  {Pradhan}, A.~K. 2012, \apj, 751, 7, \dodoi{10.1088/0004-637X/751/1/7}

\bibitem[{{Giustini} \& {Proga}(2019)}]{2019A&A...630A..94G}
{Giustini}, M., \& {Proga}, D. 2019, \aap, 630, A94,
  \dodoi{10.1051/0004-6361/201833810}

\bibitem[{{Grevesse} \& {Sauval}(1998)}]{1998SSRv...85..161G}
{Grevesse}, N., \& {Sauval}, A.~J. 1998, \ssr, 85, 161,
  \dodoi{10.1023/A:1005161325181}

\bibitem[{{Hamann} \& {Ferland}(1999)}]{1999ARA&A..37..487H}
{Hamann}, F., \& {Ferland}, G. 1999, \araa, 37, 487,
  \dodoi{10.1146/annurev.astro.37.1.487}

\bibitem[{{Holweger}(2001)}]{2001AIPC..598...23H}
{Holweger}, H. 2001, in American Institute of Physics Conference Series, Vol.
  598, Joint SOHO/ACE workshop ``Solar and Galactic Composition'', ed. R.~F.
  {Wimmer-Schweingruber}, 23--30, \dodoi{10.1063/1.1433974}

\bibitem[{{Jin} {et~al.}(2012){Jin}, {Ward}, \& {Done}}]{2012MNRAS.425..907J}
{Jin}, C., {Ward}, M., \& {Done}, C. 2012, \mnras, 425, 907,
  \dodoi{10.1111/j.1365-2966.2012.21272.x}

\bibitem[{{Kim} {et~al.}(2009){Kim}, {Stiavelli}, {Trenti}, {Pavlovsky},
  {Djorgovski}, {Scarlata}, {Stern}, {Mahabal}, {Thompson}, {Dickinson},
  {Panagia}, \& {Meylan}}]{2009ApJ...695..809K}
{Kim}, S., {Stiavelli}, M., {Trenti}, M., {et~al.} 2009, \apj, 695, 809,
  \dodoi{10.1088/0004-637X/695/2/809}

\bibitem[{{Korista} {et~al.}(1997){Korista}, {Baldwin}, {Ferland}, \&
  {Verner}}]{1997ApJS..108..401K}
{Korista}, K., {Baldwin}, J., {Ferland}, G., \& {Verner}, D. 1997, \apjs, 108,
  401, \dodoi{10.1086/312966}

\bibitem[{{Kova{\v{c}}evi{\'c}-Doj{\v{c}}inovi{\'c}} \&
  {Popovi{\'c}}(2015)}]{2015ApJS..221...35K}
{Kova{\v{c}}evi{\'c}-Doj{\v{c}}inovi{\'c}}, J., \& {Popovi{\'c}}, L.~{\v{C}}.
  2015, \apjs, 221, 35, \dodoi{10.1088/0067-0049/221/2/35}

\bibitem[{{Kurk} {et~al.}(2007){Kurk}, {Walter}, {Fan}, {Jiang}, {Riechers},
  {Rix}, {Pentericci}, {Strauss}, {Carilli}, \& {Wagner}}]{2007ApJ...669...32K}
{Kurk}, J.~D., {Walter}, F., {Fan}, X., {et~al.} 2007, \apj, 669, 32,
  \dodoi{10.1086/521596}

\bibitem[{{Kwan} \& {Krolik}(1981)}]{1981ApJ...250..478K}
{Kwan}, J., \& {Krolik}, J.~H. 1981, \apj, 250, 478, \dodoi{10.1086/159395}

\bibitem[{{Leighly} {et~al.}(2007){Leighly}, {Halpern}, {Jenkins}, \&
  {Casebeer}}]{2007ApJS..173....1L}
{Leighly}, K.~M., {Halpern}, J.~P., {Jenkins}, E.~B., \& {Casebeer}, D. 2007,
  \apjs, 173, 1, \dodoi{10.1086/519768}

\bibitem[{{Marinucci} {et~al.}(2018){Marinucci}, {Bianchi}, {Braito}, {Matt},
  {Nardini}, \& {Reeves}}]{2018MNRAS.478.5638M}
{Marinucci}, A., {Bianchi}, S., {Braito}, V., {et~al.} 2018, \mnras, 478, 5638,
  \dodoi{10.1093/mnras/sty1436}

\bibitem[{{Marziani} \& {Sulentic}(2014)}]{2014AdSpR..54.1331M}
{Marziani}, P., \& {Sulentic}, J.~W. 2014, Advances in Space Research, 54,
  1331, \dodoi{10.1016/j.asr.2013.10.007}

\bibitem[{{Marziani} {et~al.}(2001){Marziani}, {Sulentic}, {Zwitter},
  {Dultzin-Hacyan}, \& {Calvani}}]{2001ApJ...558..553M}
{Marziani}, P., {Sulentic}, J.~W., {Zwitter}, T., {Dultzin-Hacyan}, D., \&
  {Calvani}, M. 2001, \apj, 558, 553, \dodoi{10.1086/322286}

\bibitem[{{Mathews} \& {Ferland}(1987)}]{1987ApJ...323..456M}
{Mathews}, W.~G., \& {Ferland}, G.~J. 1987, \apj, 323, 456,
  \dodoi{10.1086/165843}

\bibitem[{{Matteucci} \& {Recchi}(2001)}]{2001ApJ...558..351M}
{Matteucci}, F., \& {Recchi}, S. 2001, \apj, 558, 351, \dodoi{10.1086/322472}

\bibitem[{{Mazzucchelli} {et~al.}(2017){Mazzucchelli}, {Ba{\~n}ados},
  {Venemans}, {Decarli}, {Farina}, {Walter}, {Eilers}, {Rix}, {Simcoe},
  {Stern}, {Fan}, {Schlafly}, {De Rosa}, {Hennawi}, {Chambers}, {Greiner},
  {Burgett}, {Draper}, {Kaiser}, {Kudritzki}, {Magnier}, {Metcalfe}, {Waters},
  \& {Wainscoat}}]{2017ApJ...849...91M}
{Mazzucchelli}, C., {Ba{\~n}ados}, E., {Venemans}, B.~P., {et~al.} 2017, \apj,
  849, 91, \dodoi{10.3847/1538-4357/aa9185}

\bibitem[{{Netzer}(2020)}]{2020MNRAS.494.1611N}
{Netzer}, H. 2020, \mnras, 494, 1611, \dodoi{10.1093/mnras/staa767}

\bibitem[{{Netzer} \& {Wills}(1983)}]{1983ApJ...275..445N}
{Netzer}, H., \& {Wills}, B.~J. 1983, \apj, 275, 445, \dodoi{10.1086/161545}

\bibitem[{{Osterbrock}(1977)}]{1977ApJ...215..733O}
{Osterbrock}, D.~E. 1977, \apj, 215, 733, \dodoi{10.1086/155407}

\bibitem[{{Panda} {et~al.}(2018){Panda}, {Czerny}, {Adhikari}, {Hryniewicz},
  {Wildy}, {Kuraszkiewicz}, \& {{\'S}niegowska}}]{2018ApJ...866..115P}
{Panda}, S., {Czerny}, B., {Adhikari}, T.~P., {et~al.} 2018, \apj, 866, 115,
  \dodoi{10.3847/1538-4357/aae209}

\bibitem[{{Panda} {et~al.}(2019){Panda}, {Marziani}, \&
  {Czerny}}]{2019ApJ...882...79P}
{Panda}, S., {Marziani}, P., \& {Czerny}, B. 2019, \apj, 882, 79,
  \dodoi{10.3847/1538-4357/ab3292}

\bibitem[{{Peterson}(1993)}]{1993PASP..105..247P}
{Peterson}, B.~M. 1993, \pasp, 105, 247, \dodoi{10.1086/133140}

\bibitem[{{Phillips}(1978)}]{1978ApJS...38..187P}
{Phillips}, M.~M. 1978, \apjs, 38, 187, \dodoi{10.1086/190553}

\bibitem[{{Ruff} {et~al.}(2012){Ruff}, {Floyd}, {Korista}, {Webster}, {Porter},
  \& {Ferland}}]{2012JPhCS.372a2069R}
{Ruff}, A.~J., {Floyd}, D. J.~E., {Korista}, K.~T., {et~al.} 2012, in Journal
  of Physics Conference Series, Vol. 372, Journal of Physics Conference Series,
  012069, \dodoi{10.1088/1742-6596/372/1/012069}

\bibitem[{{Sarkar} \& {Samui}(2019)}]{2019PASP..131g4101S}
{Sarkar}, A., \& {Samui}, S. 2019, \pasp, 131, 074101,
  \dodoi{10.1088/1538-3873/ab10ea}

\bibitem[{{Schindler} {et~al.}(2020){Schindler}, {Farina}, {Banados}, {Eilers},
  {Hennawi}, {Onoue}, {Venemans}, {Walter}, {Wang}, {Davies}, {Decarli}, {De
  Rosa}, {Drake}, {Fan}, {Mazzucchelli}, {Rix}, {Worseck}, \&
  {Yang}}]{2020arXiv201006902S}
{Schindler}, J.-T., {Farina}, E.~P., {Banados}, E., {et~al.} 2020, arXiv
  e-prints, arXiv:2010.06902.
\newblock \doarXiv{2010.06902}

\bibitem[{{Shen} \& {Ho}(2014)}]{2014Natur.513..210S}
{Shen}, Y., \& {Ho}, L.~C. 2014, \nat, 513, 210, \dodoi{10.1038/nature13712}

\bibitem[{{Shin} {et~al.}(2019){Shin}, {Nagao}, {Woo}, \&
  {Le}}]{2019ApJ...874...22S}
{Shin}, J., {Nagao}, T., {Woo}, J.-H., \& {Le}, H. A.~N. 2019, \apj, 874, 22,
  \dodoi{10.3847/1538-4357/ab05da}

\bibitem[{{Smyth} {et~al.}(2019){Smyth}, {Ramsbottom}, {Keenan}, {Ferland }, \&
  {Ballance}}]{2019MNRAS.483..654S}
{Smyth}, R.~T., {Ramsbottom}, C.~A., {Keenan}, F.~P., {Ferland }, G.~J., \&
  {Ballance}, C.~P. 2019, \mnras, 483, 654, \dodoi{10.1093/mnras/sty3198}

\bibitem[{{Sulentic} {et~al.}(2000){Sulentic}, {Zwitter}, {Marziani}, \&
  {Dultzin-Hacyan}}]{2000ApJ...536L...5S}
{Sulentic}, J.~W., {Zwitter}, T., {Marziani}, P., \& {Dultzin-Hacyan}, D. 2000,
  \apjl, 536, L5, \dodoi{10.1086/312717}

\bibitem[{{Tayal} \& {Zatsarinny}(2018)}]{2018PhRvA..98a2706T}
{Tayal}, S.~S., \& {Zatsarinny}, O. 2018, \pra, 98, 012706,
  \dodoi{10.1103/PhysRevA.98.012706}

\bibitem[{{Temple} {et~al.}(2020){Temple}, {Ferland}, {Rankine}, {Hewett},
  {Badnell}, {Ballance}, {Del Zanna}, \& {Dufresne}}]{2020MNRAS.496.2565T}
{Temple}, M.~J., {Ferland}, G.~J., {Rankine}, A.~L., {et~al.} 2020, \mnras,
  496, 2565, \dodoi{10.1093/mnras/staa1717}

\bibitem[{{Verner} {et~al.}(2003){Verner}, {Bruhweiler}, {Verner}, {Johansson},
  \& {Gull}}]{2003ApJ...592L..59V}
{Verner}, E., {Bruhweiler}, F., {Verner}, D., {Johansson}, S., \& {Gull}, T.
  2003, \apjl, 592, L59, \dodoi{10.1086/377571}

\bibitem[{{Verner} {et~al.}(1999){Verner}, {Verner}, {Korista}, {Ferguson},
  {Hamann}, \& {Ferland}}]{1999ApJS..120..101V}
{Verner}, E.~M., {Verner}, D.~A., {Korista}, K.~T., {et~al.} 1999, \apjs, 120,
  101, \dodoi{10.1086/313171}

\bibitem[{{V{\'e}ron-Cetty} {et~al.}(2004){V{\'e}ron-Cetty}, {Joly}, \&
  {V{\'e}ron}}]{2004A&A...417..515V}
{V{\'e}ron-Cetty}, M.~P., {Joly}, M., \& {V{\'e}ron}, P. 2004, \aap, 417, 515,
  \dodoi{10.1051/0004-6361:20035714}

\bibitem[{{Vestergaard} \& {Wilkes}(2001)}]{2001ApJS..134....1V}
{Vestergaard}, M., \& {Wilkes}, B.~J. 2001, \apjs, 134, 1,
  \dodoi{10.1086/320357}

\bibitem[{{Wang} {et~al.}(2016){Wang}, {Ferland}, {Yang}, {Wang}, \&
  {Zhang}}]{2016ApJ...824..106W}
{Wang}, T., {Ferland}, G.~J., {Yang}, C., {Wang}, H., \& {Zhang}, S. 2016,
  \apj, 824, 106, \dodoi{10.3847/0004-637X/824/2/106}

\bibitem[{{Wills} {et~al.}(1985){Wills}, {Netzer}, \&
  {Wills}}]{1985ApJ...288...94W}
{Wills}, B.~J., {Netzer}, H., \& {Wills}, D. 1985, \apj, 288, 94,
  \dodoi{10.1086/162767}

\bibitem[{{Wu} {et~al.}(2015){Wu}, {Wang}, {Fan}, {Yi}, {Zuo}, {Bian}, {Jiang},
  {McGreer}, {Wang}, {Yang}, {Yang}, {Thompson}, \&
  {Beletsky}}]{2015Natur.518..512W}
{Wu}, X.-B., {Wang}, F., {Fan}, X., {et~al.} 2015, \nat, 518, 512,
  \dodoi{10.1038/nature14241}

\end{thebibliography}
\bibliographystyle{aasjournal}

\end{document}